\documentclass[showpacs,twocolumn,aps,nofootinbib]{revtex4-1}
\usepackage[english]{babel}
\usepackage[utf8]{inputenc}
\usepackage{graphicx}
\usepackage[thinlines]{easytable}
\usepackage{epstopdf}
\usepackage{amsmath, amssymb}
\usepackage{dcolumn}
\usepackage[mathscr]{euscript}
\usepackage{caption}
\usepackage{subcaption} 
\usepackage{amsmath,amsfonts,amsthm,bm}
\usepackage{mathtools}
\usepackage{float}
\usepackage{color}
\usepackage{braket}
\usepackage{wasysym}

\DeclareMathOperator*{\SumInt}{%
	\mathchoice%
	{\ooalign{$\displaystyle\sum$\cr\hidewidth$\displaystyle\int$\hidewidth\cr}}
	{\ooalign{\raisebox{.14\height}{\scalebox{.7}{$\textstyle\sum$}}\cr\hidewidth$\textstyle\int$\hidewidth\cr}}
	{\ooalign{\raisebox{.2\height}{\scalebox{.6}{$\scriptstyle\sum$}}\cr$\scriptstyle\int$\cr}}
	{\ooalign{\raisebox{.2\height}{\scalebox{.6}{$\scriptstyle\sum$}}\cr$\scriptstyle\int$\cr}}
}

\newcommand{\be}{\begin{equation}}
\newcommand{\ee}{\end{equation}}
\newcommand{\bq}{\begin{eqnarray}}
\newcommand{\eq}{\end{eqnarray}}

\begin{document}
\renewcommand{\figurename}{FIG.}

\title{Entanglement and scattering in quantum electrodynamics: S-matrix information from an entangled spectator particle}

\author{Juan D. Fonseca$^a$}\email[]{juan.fonseca@ufabc.edu.br}
\author{B. Hiller$^b$}\email[]{brigitte@fis.uc.pt}
\author{J. B. Araujo $^d$}\email[]{jonas.bastosdearaujo@colibritd.com}
\author{I. G. da Paz$^c$}\email[]{irismar@ufpi.edu.br}
\author{M. Sampaio $^a$}\email[]{marcos.sampaio@ufabc.edu.br}

\affiliation{$^{a}$  CCNH, Universidade Federal do ABC,  09210-580 , Santo Andr\'e - SP, Brazil}
\affiliation{$^{b}$  CFisUC - Department of Physics, University of Coimbra, 3004-516 Coimbra, Portugal}
\affiliation{$^c$ Universidade Federal do Piau\'{\i}, Departamento de F\'{\i}sica, 64049-550, Teresina, PI, Brazil}
\affiliation{$^d$ ColibrITD Quantum R\&D, 1-7 Cours de Valmy, 92800, Puteaux, France.}

\begin{abstract}
\noindent

We consider a general quantum field relativistic scattering involving two half spin fermions, $A$ and $B$, which are initially entangled  with another fermion $C$ that does not participate in the scattering dynamics. We construct general expressions for the reduced spin matrices for the out-state considering a general tripartite spin-entangled state. In particular we study an inelastic QED process at tree-level, namely $e^-e^+\rightarrow \mu^- \mu^+$ and a half spin fermion $C$ as a spectator particle which can be entangled to the $AB$ system in the following ways: W state, GHZ state, $|\text{A}^\alpha \rangle \otimes |\Psi^{\pm} \rangle_{\text{BC}}$ and $|\text{A}^\alpha \rangle \otimes |\Phi^{\pm} \rangle_{\text{BC}}$, where $\{|\Psi^{\pm} \rangle,|\Phi^{\pm} \rangle\}$ are the Bell basis states and $|\text{A}^\alpha \rangle$ is a spin superposition state of system $A$. We calculate the von-Neumann entropy variation before and after the scattering for the particle $C$ and show that spin measurements in $C$ contain numerical information about the total cross section of the process. We compare the initial states W and GHZ as well as study the role played by the parameter $\alpha$ in the evaluation of the entropy variations and the cross section encoded in the spectator particle.

\end{abstract}

\maketitle

\section{Introduction}

The counter-intuitive properties of entanglement were formalized in the famous 1935 paper of Einstein, Podolsky and Rosen \cite{EPR}. This work paved the way to study theoretically and experimentally the behaviour of global states of composite systems. Rather than an incomplete description of physical reality or the lack of extra variables to comply with local realism, the violation of Bell's inequalities \cite{BELL} and its variants crowns quantum mechanics as a non-local and complete theory without contradicting relativistic  causality \cite{GRANGIER}.
The experimental verification of quantum non-locality became possible only in the early eighties through an ingenious
experiment in Aspect's laboratory in France \cite{ASPECT}. Later on, possible locality and detection loopholes were discarded by state-of-the-art technological refinements such as in the experiment involving entangled electron spins separated by around 1.3 km \cite{LOOPHOLE1} as well as atoms separated by 398 m via measurements of the atomic spin states \cite{LOOPHOLE2}.

A growing interest in relativistic aspects of quantum entanglement (see \cite{RQIREVIEW} for a review) is reflected in many works that appeared at the turn of the $21^{st}$ century. Relativistic quantum information sets out to answer, amongst other questions,  how quantum information processing is affected by relativistic motion  both in inertial and non-inertial frames \cite{FRIIS}. In Ref. \cite{BERTLMANN}, an entanglement measure between different partitions of a composite momentum-spin state with two half-spin particles was studied in regard to Lorentz boosts. Also, Lorentz frame dependence of Bell inequalities as a function of  rapidities
\cite{AHN}, as well as on specific choices of boost directions \cite{BERTLMANN} have been investigated.  Relativistic aspects of quantum information is motivated by the possibility to  exploit relativistic
effects to enhance current quantum technologies through new
relativistic quantum protocols \cite{RQIIF}. Non-inertial effects on entanglement generation  are relevant to precision measurements
in cosmology \cite{GW} and the technology involving photonic interferometric devices \cite{PID}. Entanglement generation throughout particle production in expanding spacetimes \cite{BALL} and the deleterious effect of acceleration on  entanglement of quantum field modes inside cavities
\cite{ALICE}, \cite{LOUKO} have also been theoretically studied.

Many contributions have appeared to study quantum entanglement in spin and momentum degrees of freedom as a result of particle collisions in both non-relavistic  \cite{FEDER, GHANBARI, HARSHMAN4, HARSHMAN1, HARSHMAN2, HARSHMAN3, HARSHMAN5, LAMATALEON, KOUZAKOV} and relativistic formulations \cite{PACHOS, MISHIMA, LAMATASOLANO, WANG, BUSCEMI} or involving other internal symmetries \cite{102001, 168581, 2150205}. The effects of initial state quantum correlations on the cross-section was considered in Refs. \cite{SEKI1, SEKI2, SEKI3, Faleiro}.  In Ref. \cite{FAN1}, the authors use a QED inelastic scattering to show that the variation of entanglement entropy between a pair $e^+e^-$ in the initial state and final state composed of a pair $\mu^+\mu^-$ is proportional to  the total cross section. The same scattering process was used in \cite{FAN2} to study the change of entanglement entropy as seen from different Lorentz observers. Because such a quantity is proportional to the cross-section, it was explicitly verified through appropriate Wigner rotations that it is Lorentz invariant whilst the correlations involving only spin degrees of freedoms is not. Bell's inequality violation in QED was verified in the process
$2\gamma\rightarrow e^+e^-$ \cite{YONGRAM} and its experimental measurement was contemplated. Along the same lines, it has been shown that  the initial state quantum correlations may have a direct effect on the cross section. For instance, the $\gamma \gamma$ QED scattering realized at one loop order through the exchange of virtual electron-positron pairs is known to have a tiny cross-section ($\approx 10^{-24} \, \text{m}^2$ for $\lambda= 30$pm \cite{INADA}). In \cite{RATZELQED}, the authors related  the differential cross section of photon-photon scattering to  the degree of polarization entanglement of the two-photon initial state. As a function of the scattering angle, the scattering becomes stronger (weaker) for the symmetric (antisymmetric) Bell state than for a factorized initial state for photons (see also \cite{RATZELQG}).
 In \cite{CABAN1},\cite{CABAN2} relativistic spin correlations for initially polarized electrons scattered off by polarized and unpolarized targets were calculated
showing that the Clauser-Horne-Shimony-Holt (CHSH) inequality can be violated for relativistic energies when both scattering electrons are highly polarized in some specific directions. The same system was used in \cite{TSURIKOV} to derive a spin correlation tensor for the two electron state.

The quantum field theoretical description of quantum information in $S$-matrix calculations opens new venues for questions related to the amount of information drawn off by soft radiation emitted by asymptotic states representing charged particles. The subject becomes even more interesting as finite observables can only be obtained after taking into account the emission of soft photons as
stated by the Bloch and Nordsieck (BN) and the Kinoshita, Lee and
Nauenberg (KLN) theorem \cite{KLN}.  The loss of coherence stemming from the integration of infrared degrees of freedom has been studied for instance in \cite{GOMEZ}. Moreover, in \cite{CARNEY} it was claimed that unobserved soft photons decohere momentum superpositions of charged particles and that the dressed state formalism can lead to more sensible physical results\cite{TOMARAS}.

Finally, the QED inelastic process $e^+e^-\rightarrow \mu^+\mu^-$ watched by a spectator particle, entangled with one of the particles in the in-state has been considered in \cite{PRD2019}. It was claimed that partial information about the dynamics and the total cross-section of particles $A$, $B$ could be obtained from spin measurements on the spectator particle $C$. In the present contribution we provide a more complete analysis as the one presented in \cite{PRD2019} by considering a general tripartite spin $1/2$-state to study how measurements of spin on the spectator particle can be optimized with respect to initial state parameters to provide indirect information of the cross section.

This contribution is organized as follows:
 In Sec. \ref{general}, we set out our notation , including the \textit{in-}state proposed with general spin configuration and, using the scattering matrix formalism, we construct  the \textit{out-}state. In Sec. \ref{total}, the final reduced density
matrix of spectator particle $C$ is calculated which is concretely realized in Sec. \ref{inelastic} for the  QED process $e^{+}e^{-}\rightarrow\mu^{+}\mu^{-}$. A discussion of our results and perspectives are left to Sec. \ref{conclusions}.

\section{In-state and 2-particle scattering}
\label{general}

Consider a general tripartite spin $1/2$-state
\begin{eqnarray}
	\ket{\mathscr{S}} &=& c_{1}\ket{\uparrow\uparrow\uparrow}+c_{2}\ket{\uparrow\uparrow\downarrow}+c_{3}\ket{\uparrow\downarrow\uparrow} \nonumber \\ &+&
	c_{4}\ket{\uparrow\downarrow\downarrow}+c_{5}\ket{\downarrow\uparrow\uparrow}+c_{6}\ket{\downarrow\uparrow\downarrow} \nonumber \\ &+&
	c_{7}\ket{\downarrow\downarrow\uparrow}+c_{8}\ket{\downarrow\downarrow\downarrow},
	\label{estadogeneralinicial}
\end{eqnarray}
in which Alice, Bob and Claire state $|s_\text{A} s_\text{B} s_\text{C}\rangle \equiv |s_\text{A}\rangle \otimes |s_\text{B}\rangle \otimes |s_\text{C}\rangle$ is a three-particle spin-state constrained by $\sum_{j=1}^{8}|c_{j}|^{2}=1$. For specific sets of values of $c_{j}$, a Greenberger-Horne-Zeilinger ($c_1=c_8 =1/\sqrt{2}$ and $c_{i} = 0$ otherwise)
\begin{equation}
\ket{\textup{GHZ}}\equiv\frac{1}{\sqrt{2}}[\ket{\downarrow\downarrow\downarrow}+\ket{\uparrow\uparrow\uparrow}],	
\end{equation}
or a Wolfgang D\"ur-Vidal-Cirac  ($c_4=c_6=c_7=1/\sqrt{3}$ and $c_i=0$ otherwise) 
\begin{equation}
\ket{\textup{W}}\equiv\frac{1}{\sqrt{3}}[\ket{\downarrow\downarrow\uparrow}+\ket{\downarrow\uparrow\downarrow}+\ket{\uparrow\downarrow\downarrow}]	
\end{equation} 
state  is recovered \cite{vidal}. GHZ and W three-qubit states are well studied inequivalent classes of states under local operations and classical communications. Whilst entanglement in the W state is robust
against one particle loss, the GHZ state is maximally entangled as it is reduced to a product of two qubits. They have been applied in various contexts, for instance in algorithms for the generation of entangled GHZ and W states of up to 16 qubits useful in quantum networks \cite{CRUZ}, and in the study of quantum teleportation through noisy channels \cite{JUNG}. Of particular interest to our work is the product state formed by a spin superposition of Alice's spin state and a general entangled Bob and Claire state, which we divide into two special cases:
\begin{itemize}
\item $c_{1}=c_{4}=c_{5}=c_{8}=0, c_{2}=\cos\alpha\cos\eta, c_{3}=\cos\alpha\sin\eta, c_{6}=\sin\alpha\cos\eta, c_{7}=\sin\alpha\sin\eta$:
\begin{eqnarray}
\ket{\textup{A}^\alpha}\otimes\ket{\Psi^{\eta}}&=&[\cos\alpha\ket{\uparrow}+\sin\alpha\ket{\downarrow}]\nonumber \\ &\otimes&[\cos\eta\ket{\uparrow\downarrow}+\sin\eta\ket{\downarrow\uparrow}].
\label{APSI}
\end{eqnarray}
Particularly, if $\eta=\pi/4$, we get $\ket{\textup{A}^\alpha}\otimes\ket{\Psi^{+}}$ where $\ket{\Psi^{+}}=(2)^{-1/2}[\ket{\uparrow\downarrow}+\ket{\downarrow\uparrow}]$; if  $\eta=3\pi/4$, we get $\ket{\textup{A}^\alpha}\otimes\ket{\Psi^{-}}$ where $\ket{\Psi^{-}}=(2)^{-1/2}[\ket{\downarrow\uparrow}-\ket{\uparrow\downarrow}]$.
\item $c_{2}=c_{3}=c_{6}=c_{7}=0, c_{1}=\cos\alpha\cos\eta, c_{4}=\cos\alpha\sin\eta, c_{5}=\sin\alpha\cos\eta, c_{8}=\sin\alpha\sin\eta$:
\begin{eqnarray}
\ket{\textup{A}^\alpha}\otimes\ket{\Phi^{\eta}}&=&[\cos\alpha\ket{\uparrow}+\sin\alpha\ket{\downarrow}]\nonumber \\
&\otimes&[\cos\eta\ket{\uparrow\uparrow}+\sin\eta\ket{\downarrow\downarrow}].
\label{APHI}
\end{eqnarray}
Particularly, if $\eta=\pi/4$, we get $\ket{\textup{A}^\alpha}\otimes\ket{\Phi^{+}}$ where $\ket{\Phi^{+}}=(2)^{-1/2}[\ket{\uparrow\uparrow}+\ket{\downarrow\downarrow}]$;  if $\eta=3\pi/4$, we get $\ket{\textup{A}^\alpha}\otimes\ket{\Phi^{-}}$ where $\ket{\Phi^{-}}=(2)^{-1/2}[\ket{\downarrow\downarrow}-\ket{\uparrow\uparrow}]$.
\end{itemize}
The states $\ket{\Psi^\pm}$ and $\ket{\Phi^\pm}$ are the Bell basis  \cite{bellstates} for Bob and Claire spins. Remarkably, we shall verify in this work that the parameter $\alpha$ plays a crucial role in the quantum coherences developed by the spectator Claire which contain information on Alice and Bob dynamics.

 As usual, one describes asymptotic particle states in the distant past and future ($t\rightarrow \pm \infty$)
as wave packets which are well separated in position space and narrowly peaked in momentum space. Here, we take them to be plane waves with definite momenta. The unitary scattering matrix $\boldsymbol{\widehat{\mathcal{S}}}$ is defined as the time evolution operator in the interaction
picture in terms of asymptotic states. Therefore for an initially factorized state, we have
\bq
\ket{\psi_{\text{in}}^{\text{fact}}} &=& \bigotimes_{j=1}^n\ket{\textbf{\textup{p}}_{\textup{i}}^j,s_{\textup{i}}^j},\nonumber \\
\ket{\psi_{\textup{out}}} &=& \mathscr{N}^{-1}\bigg[\bigotimes_{k=1}^{m}\mathcal{\widehat{I}}^{(k)}_{\textup{f}}\bigg]\boldsymbol{\widehat{ \mathcal{S}}}\ket{\psi_{\textup{in}}^{\text{fact}}},
\eq
for $n(m)$ initial (final) particles, with
\begin{equation}
\mathcal{\widehat{I}}_{\textup{f}} =\SumInt_{\bf{p}_{f},s_{f}}\ket{\textbf{\textup{p}}_{\textup{f}},s_{\textup{f}}}\bra{\textbf{\textup{p}}_{\textup{f}},s_{\textup{f}}},\label{completeness}
\end{equation}
 in which for short $\int_{\textbf{\textup{p}}_{\textup{f}}}\equiv[(2\pi)^{3}2E_{\textbf{\textup{p}}_{\textup{f}}}]^{-1}d^{3}\textbf{\textup{p}}_{\textup{f}}$ is the Lorentz invariant measure over final momentum and $\sum_{s_{\textup{f}}}$ is the sum over final spin possibilities. The factor $\mathscr{N}^{-1}$ is such that $\ket{\psi_{\textup{out}}}$ is normalized. It is useful to define the transition matrix $\boldsymbol{\widehat{\mathcal{T}}}$ such that $\boldsymbol{\widehat{\mathcal{S}}}=\boldsymbol{\widehat{\mathcal{I}}}+i\boldsymbol{\widehat{\mathcal{T}}}$ which accounts for the non-trivial of scattering amplitudes and is related to the transition amplitude $\mathcal{M}$ as given  by Feynman diagrams as
\begin{eqnarray}
\braket{\textbf{\textup{p}}_{\textup{f}}^{(1)},s_{\textup{f}}^{(1)};\textbf{\textup{p}}_{\textup{f}}^{(2)},s_{\textup{f}}^{(2)}|i\boldsymbol{\widehat{\mathcal{T}}}|\textbf{\textup{p}}_{\textup{i}}^{(1)},s_{\textup{i}}^{(1)};\textbf{\textup{p}}_{\textup{i}}^{(2)},s_{\textup{i}}^{(2)}}&&= \nonumber \\  i(2\pi)^{4}\delta^{4}(p_{\text{i}}^{(1)}+p_{\text{i}}^{(2)}-p_{\text{f}}^{(1)}-p_{\text{f}}^{(2)})\mathcal{M}_{\text{i}\rightarrow\text{f}},&&\label{ToperatorM}
\end{eqnarray}
for $n=m=2$, and
\begin{equation}
\mathcal{M}_{\text{i}\rightarrow\text{f}}=\mathcal{M}^{\textbf{\text{p}}_{\text{i}}^{(1)},\textbf{\text{p}}_{\text{i}}^{(2)}}_{\textbf{\text{p}}_{\text{f}}^{(1)},\textbf{\text{p}}_{\text{f}}^{(2)}}(s_{\text{i}}^{(1)}s_{\text{i}}^{(2)}\rightarrow s_{\text{f}}^{(1)}s_{\text{f}}^{(2)}).\label{Mvectors}
\end{equation}

We study the 2-particle scattering of the tripartite \textit{in-} spin-$1/2$ state which we call Alice, Bob and Claire states, factorized in the momentum variables and whose spin is described by the general state in equation (\ref{estadogeneralinicial}), namely
\begin{equation}
\ket{\psi_{\text{in}}}=\bigg[\bigotimes_{j=1}^{3}\ket{\textup{\textbf{p}}_{j}}\bigg]\otimes\ket{\mathscr{S}},
\label{eqn:estado-inicial}
\end{equation}
where, for shorthand notation, we set $\ket{\textup{\textbf{p}}_{1}}\equiv\ket{\textup{\textbf{p}}_{\textup{i}}^{(1)}}$, $\ket{\textup{\textbf{p}}_{2}}\equiv\ket{\textup{\textbf{p}}_{\textup{i}}^{(2)}}$ and $\ket{\textup{\textbf{q}}}\equiv\ket{\textup{\textbf{p}}_{\textup{i}}^{(3)}}$. Then the \textit{out-}state reads
\begin{eqnarray}
\ket{\psi_{\text{out}}}&=&\ket{\psi_{\text{in}}}+i\SumInt_{ \textup{\textbf{p}}_{3},\textup{\textbf{p}}_{4},r,s}(2\pi)\delta(E_{\textbf{\text{p}}_{1}}+E_{\textbf{\text{p}}_{2}}-E_{\textbf{\text{p}}_{3}}-E_{\textbf{\text{p}}_{4}})\times\nonumber \\
& &\times(2\pi)^{3}\delta^{3}(\textbf{\text{p}}_{1}+
\textbf{\text{p}}_{2}-\textbf{\text{p}}_{3}-\textbf{\text{p}}_{4})
\Big[c_{1}\mathcal{M}(\uparrow\uparrow\rightarrow rs)\ket{\textup{\textbf{q}},\uparrow} \nonumber \\ & &+c_{2}\mathcal{M}(\uparrow\uparrow\rightarrow rs)\ket{\textup{\textbf{q}},\downarrow} +c_{3}\mathcal{M}(\uparrow\downarrow\rightarrow rs)\ket{\textup{\textbf{q}},\uparrow} \nonumber \\
& &+c_{4}\mathcal{M}(\uparrow\downarrow\rightarrow rs)\ket{\textup{\textbf{q}},\downarrow}+c_{5}\mathcal{M}(\downarrow\uparrow\rightarrow rs)\ket{\textup{\textbf{q}},\uparrow}
\nonumber \\ 
& &+c_{6}\mathcal{M}(\downarrow\uparrow\rightarrow rs)\ket{\textup{\textbf{q}},\downarrow}
+c_{7}\mathcal{M}(\downarrow\downarrow\rightarrow rs)\ket{\textup{\textbf{q}},\uparrow} \nonumber \\
& &+c_{8}\mathcal{M}(\downarrow\downarrow\rightarrow rs)\ket{\textup{\textbf{q}},\downarrow}\Big]\otimes\ket{\textbf{\text{p}}_{3},r}\otimes\ket{\textbf{\text{p}}_{4},s} \nonumber \\
& & \equiv \ket{\psi_{\text{in}}} + \ket{\psi_{\text{trans}}},
\label{finalsimplifi}
\end{eqnarray}
where the subscript ``trans'' stands for transition, the non-trivial part of the $out$-state and we omitted the momentum indices  $\textbf{\text{p}}_{1},\textbf{\text{p}}_{2},\textbf{\text{p}}_{3},\textbf{\text{p}}_{4}$ in the $\mathcal{M}$'s. It is noteworthy that in the case of an inelastic QED scattering such as $e^{+}e^{-}\rightarrow \mu^{+}\mu^{-}$, the amplitude $\mathcal{M}$ has the \textit{s}-channel only. Thus we call the particle states $\ket{\textup{\textbf{p}}_{1},\uparrow(\downarrow)}$ and $\ket{\textup{\textbf{p}}_{3},r}$ as system $A$, whereas the antiparticle states $\ket{\textup{\textbf{p}}_{2},\uparrow(\downarrow)}$ and $\ket{\textup{\textbf{p}}_{4},s}$  are called system $B$. The spectator particle $\ket{\textup{\textbf{q}},\uparrow(\downarrow)}$, 
is called system $C$. The total density matrix is given by 
\be
\rho_{\text{out}}= \mathcal{N}^{-1}\ket{\psi_{\text{out}}}\bra{\psi_{\text{out}}},
\ee
and the reduced density matrices of a system are obtained by tracing out the other systems, with the trace defined by
$\text{Tr}[\rho]=\sum_{\sigma}\int_{\textbf{\text{k}}}\bra{\textbf{\text{k}},\sigma}\rho\ket{\textbf{\text{k}},\sigma}$, to yield
\bq
 \rho_{\text{A}}&=&\text{Tr}_{\text{B}}\text{Tr}_{\text{C}}[\rho_{\text{out}}],\nonumber \\
\rho_{\text{B}}&=&\text{Tr}_{\text{A}}\text{Tr}_{\text{C}}[\rho_{\text{out}}],\nonumber \\
\rho_{\text{C}}&=&\text{Tr}_{\text{A}}\text{Tr}_{\text{B}}[\rho_{\text{out}}].
\label{matricesreducidas}
\eq
The density matrix normalization $\mathcal{N}$, complying with $\text{Tr}[\rho_{\text{out}}]=1$, is  $\mathcal{N}=\text{Tr}_{\text{C}}\text{Tr}_{\text{A}}\text{Tr}_{\text{B}}[\rho_{\text{out}}]$. The ordering of the traces is immaterial but we start off tracing out systems $A$ and $B$ as given in equation (\ref{matricesreducidas}) since we are interested in evaluating expectation values on the system $C$. Also, it is useful to write 
\begin{equation}
\mathcal{N}=\mathcal{N}_{\text{in}}+\mathcal{N}_{\text{cross}}+\mathcal{N}_{\text{trans}},
\end{equation}
with
\bq
\mathcal{N}_{\text{in}}&=&\text{Tr}_{\text{C}}\text{Tr}_{\text{A}}\text{Tr}_{\text{B}}[\ket{\psi_{\text{in}}}\bra{\psi_{\text{in}}}], \nonumber\\
\mathcal{N}_{\text{cross}}&=&\text{Tr}_{\text{C}}\text{Tr}_{\text{A}}\text{Tr}_{\text{B}}[\ket{\psi_{\text{in}}}\bra{\psi_{\text{trans}}}+\ket{\psi_{\text{trans}}}\bra{\psi_{\text{in}}}],\nonumber\\
\mathcal{N}_{\text{trans}}&=&\text{Tr}_{\text{C}}\text{Tr}_{\text{A}}\text{Tr}_{\text{B}}[\ket{\psi_{\text{trans}}}\bra{\psi_{\text{trans}}}].
\label{matricesreducidasN}
\eq
In the trace operations, we use a Lorentz invariant particle spin-momentum state scalar product $\braket{\textbf{\text{k}},\sigma|\textbf{\text{p}},s}=(2\pi)^{3}2E_{\textbf{\text{p}}}\delta^{3}(\textbf{\text{k}}-\textbf{\text{p}})\delta_{\sigma,s}$, such that
\begin{align}
\SumInt_{\bf{k},\sigma}\frac{1}{2E_{\textbf{\text{k}}}}\braket{\textbf{\text{k}},\sigma|\textbf{\text{p}},s}\braket{\textbf{\text{p}}',s'|\textbf{\text{k}},\sigma}=2E_{\textbf{\text{p}}}(2\pi)^{3}\delta^{3}(\textbf{\text{p}}'-\textbf{\text{p}})\delta_{s',s}.
\label{traza}
\end{align}
Since $\sum_{j=1}^{8}c_{j}^{2}=1$ in the initial state
(\ref{eqn:estado-inicial}), and using
equation (\ref{traza}), we get the dimensionless quantity 
\begin{equation}
\mathcal{N}_{\text{in}} = 8E_{\textbf{\text{q}}}E^{2}V^{3},   
\end{equation}
where $V = (2 \pi)^3 \delta^3 ({\boldsymbol\epsilon})$ with ${\boldsymbol\epsilon} \rightarrow \bf{0}$ is the space volume, $E$ the energy of colliding particles ($A,B$) in the centre-of-mass frame, and $E_{\textbf{\text{q}}}$ the energy of the spectator particle C, whereas  $\mathcal{N}_{\text{cross}}$ can be easily shown to give zero. As for  $\mathcal{N}_{\text{trans}}$, using that the spin-$1/2$ state in equation (\ref{estadogeneralinicial}) is diagonal in the $z$-direction, which is also the direction of the initial momenta before collision, as well as energy momentum conservation and spherical symmetry of the final state momenta, we can simplify the integrand and obtain 
\bq
\mathcal{N} &=& \mathcal{N}_{\text{in}}+\mathcal{N}_{\text{trans}}
\nonumber \\
&=&8E_{\textbf{\text{q}}}E^{2}V^{3}+\frac{E_{\textbf{\text{q}}}TV^{2}|\textbf{\text{p}}_{3}|}{16\pi^{2}E}\SumInt_{r,s} d\Omega\Lambda(r,s),\label{explinorma}
\eq
as the total normalization. In the equation above, $\Lambda(r,s)$ stands for
\bq
\Lambda(r,s)&=&(c_{1}^{2}+c_{2}^{2}+c_{7}^{2}+c_{8}^{2})|\mathcal{M}_{cm}(\uparrow\uparrow\rightarrow rs)|^{2} \nonumber\\
&+& (c_{3}^{2}+c_{4}^{2}+c_{5}^{2}+c_{6}^{2})|\mathcal{M}_{cm}(\uparrow\downarrow\rightarrow rs)|^{2}\\
&+&2(c_{3}c_{5}+c_{4}c_{6})\mathcal{M}_{cm}(\uparrow\downarrow\rightarrow rs) \mathcal{\bar{M}}_{cm}(\downarrow\uparrow\rightarrow rs), \nonumber\label{lambda}
\eq
where, in the centre-of-mass frame, $\mathcal{M}_{cm}$ represents $\mathcal{M}^{\bf{p}_1,-\bf{p}_1}_{\bf{p}_3,-\bf{p}_3}$ evaluated at $|\textbf{\text{p}}_{3}|=\sqrt{E^{2}-m_{3}^{2}}$ and the infinitesimal solid angle is $d\Omega = \sin \theta d \theta d \phi$, with $\theta$ being the scattering angle. A barred quantity means Hermitian conjugation and $T$ is the ``time duration'' such that $2 \pi \delta (E_\text{f} - E_\text{i}) =\lim_{T\rightarrow \infty} \int_{-T/2}^{+T/2} \, dt \, \exp[i (E_\text{f} - E_\text{i}) t]$.

\section{Total spin density matrix of system $\boldsymbol{C}$}\label{total}

In order to extract information about the scattering from system $C$, we need to evaluate its reduced density matrix $\rho_\text{C}$ from equation (\ref{finalsimplifi}). By firstly tracing over system $B$, we get only two contributions
\begin{equation}
	\rho_{\text{AC}} = \frac{1}{\mathcal{N}}\Big(\rho_{\text{AC(in)}} + \rho_{\text{AC(trans)}}\Big),
	\label{rhoAC}
\end{equation}
since the crossed term vanishes, $\mathcal{N}$ is given by (\ref{explinorma}) and 
\begin{equation}
	\rho_{\text{AC(in)}}=8E_{\textbf{\text{q}}}E^{2}V^{3} \times
\end{equation}
\vskip-0.7cm
\begin{equation}\times
	\begin{pmatrix}
		c_{1}^{2}+c_{3}^{2} & c_{1}c_{2}+c_{3}c_{4} & c_{1}c_{5}+c_{3}c_{7} &  c_{1}c_{6}+c_{3}c_{8} \\\\
		c_{1}c_{2}+c_{3}c_{4} & c_{2}^{2}+c_{4}^{2} & c_{2}c_{5}+c_{4}c_{7} &  c_{2}c_{6}+c_{4}c_{8} \\\\
		c_{1}c_{5}+c_{3}c_{7} & c_{2}c_{5}+c_{4}c_{7} & c_{5}^{2}+c_{7}^{2} &  c_{5}c_{6}+c_{7}c_{8} \\\\
		c_{1}c_{6}+c_{3}c_{8} & c_{2}c_{6}+c_{4}c_{8} & c_{5}c_{6}+c_{7}c_{8} &  c_{6}^{2}+c_{8}^{2} 
	\end{pmatrix},\label{ACin}\nonumber
\end{equation}
and
\begin{equation}
	\rho_{\text{AC(trans)}}=\frac{E_{\textbf{\text{q}}}TV^{2}|\textbf{\text{p}}_{3}|}{16\pi^{2}E}\SumInt_{s} d\Omega \times 
\end{equation}
\vskip-0.5cm
\begin{equation*}\times
	\begin{pmatrix}
		\Lambda^{1,1}_{\text{AC}}(s) & \Lambda^{1,2}_{\text{AC}}(s) & \Lambda^{1,3}_{\text{AC}}(s) & \Lambda^{1,4}_{\text{AC}}(s)\\\\ 
		\Lambda^{1,2}_{\text{AC}}(s) & \Lambda^{2,2}_{\text{AC}}(s) & \Lambda^{2,3}_{\text{AC}}(s) & \Lambda^{2,4}_{\text{AC}}(s)\\\\
		\Lambda^{1,3}_{\text{AC}}(s) & \Lambda^{2,3}_{\text{AC}}(s) & \Lambda^{3,3}_{\text{AC}}(s) & \Lambda^{3,4}_{\text{AC}}(s)\\\\
		\Lambda^{1,4}_{\text{AC}}(s) & \Lambda^{2,4}_{\text{AC}}(s) & \Lambda^{3,4}_{\text{AC}}(s) & \Lambda^{4,4}_{\text{AC}}(s)
	\end{pmatrix};\label{ACex}
\end{equation*}
where the matrix elements $\Lambda^{i,j}_{\text{AC}}(s)$ are given by:
\begin{eqnarray}
	\Lambda^{1,1}_{\text{AC}}(s)&=&c_{1}^{2}|\mathcal{M}_{cm}(\uparrow\uparrow\rightarrow\uparrow s)|^{2}+c_{3}^{2}|\mathcal{M}_{cm}(\uparrow\downarrow\rightarrow\uparrow s)|^{2}\nonumber\\
	&+& c_{5}^{2}|\mathcal{M}_{cm}(\downarrow\uparrow\rightarrow\uparrow s)|^{2}+
	c_{7}^{2}|\mathcal{M}(\downarrow\downarrow\rightarrow\uparrow s)|^{2}\nonumber\\
	&+&
	2c_{3}c_{5}\mathcal{M}_{cm}(\uparrow\downarrow\rightarrow\uparrow s)\mathcal{\bar{M}}_{cm}(\downarrow\uparrow\rightarrow\uparrow s),\nonumber\\
	\Lambda^{2,2}_{\text{AC}}(s)&=&c_{2}^{2}|\mathcal{M}_{cm}(\uparrow\uparrow\rightarrow\uparrow s)|^{2}+c_{4}^{2}|\mathcal{M}_{cm}(\uparrow\downarrow\rightarrow\uparrow s)|^{2} \nonumber \\ &+&c_{6}^{2}|\mathcal{M}_{cm}(\downarrow\uparrow\rightarrow\uparrow s)|^{2}+c_{8}^{2}|\mathcal{M}_{cm}(\downarrow\downarrow\rightarrow\uparrow s)|^{2} \nonumber\\
	&+&2c_{4}c_{6}\mathcal{M}_{cm}(\uparrow\downarrow\rightarrow\uparrow s)\mathcal{\bar{M}}_{cm}(\downarrow\uparrow\rightarrow\uparrow s)\nonumber,\\
	\Lambda^{3,3}_{\text{AC}}(s)&=&c_{1}^{2}|\mathcal{M}_{cm}(\uparrow\uparrow\rightarrow\downarrow s)|^{2}+c_{3}^{2}|\mathcal{M}_{cm}(\uparrow\downarrow\rightarrow\downarrow s)|^{2} \nonumber\\ &+&c_{5}^{2}|\mathcal{M}_{cm}(\downarrow\uparrow\rightarrow\downarrow s)|^{2}+c_{7}^{2}|\mathcal{M}_{cm}(\downarrow\downarrow\rightarrow\downarrow s)|^{2}\nonumber\\
	&+&2c_{3}c_{5}\mathcal{M}_{cm}(\uparrow\downarrow\rightarrow\downarrow s)\mathcal{\bar{M}}_{cm}(\downarrow\uparrow\rightarrow\downarrow s),\nonumber\\
	\Lambda^{4,4}_{\text{AC}}(s)&=&c_{2}^{2}|\mathcal{M}_{cm}(\uparrow\uparrow\rightarrow\downarrow s)|^{2}+c_{4}^{2}|\mathcal{M}_{cm}(\uparrow\downarrow\rightarrow\downarrow s)|^{2} \nonumber\\ &+&c_{6}^{2}|\mathcal{M}_{cm}(\downarrow\uparrow\rightarrow\downarrow s)|^{2}+c_{8}^{2}|\mathcal{M}_{cm}(\downarrow\downarrow\rightarrow\downarrow s)|^{2}\nonumber\\
	&+&2c_{4}c_{6}\mathcal{M}_{cm}(\uparrow\downarrow\rightarrow\downarrow s)\mathcal{\bar{M}}_{cm}(\downarrow\uparrow\rightarrow\downarrow s),\nonumber\\
	\Lambda^{1,2}_{\text{AC}}(s)&=&c_{1}c_{2}|\mathcal{M}_{cm}(\uparrow\uparrow\rightarrow\uparrow s)|^{2}+c_{3}c_{4}|\mathcal{M}_{cm}(\uparrow\downarrow\rightarrow\uparrow s)|^{2} \nonumber \\ &+&c_{5}c_{6}|\mathcal{M}_{cm}(\downarrow\uparrow\rightarrow\uparrow s)|^{2}+c_{7}c_{8}|\mathcal{M}_{cm}(\downarrow\downarrow\rightarrow\uparrow s)|^{2}
	\nonumber\\
	&+&(c_{3}c_{6}+c_{4}c_{5})\mathcal{M}_{cm}(\uparrow\downarrow\rightarrow\uparrow s)\mathcal{\bar{M}}_{cm}(\downarrow\uparrow\rightarrow\uparrow s),\nonumber\\
	\Lambda^{3,4}_{\text{AC}}(s)&=&c_{1}c_{2}|\mathcal{M}_{cm}(\uparrow\uparrow\rightarrow\downarrow s)|^{2}+c_{3}c_{4}|\mathcal{M}_{cm}(\uparrow\downarrow\rightarrow\downarrow s)|^{2}\nonumber\\
	&+&c_{5}c_{6}|\mathcal{M}_{cm}(\downarrow\uparrow\rightarrow\downarrow s)|^{2}+c_{7}c_{8}|\mathcal{M}_{cm}(\downarrow\downarrow\rightarrow\downarrow s)|^{2}\nonumber\\
	&+&(c_{3}c_{6}+c_{4}c_{5})\mathcal{M}_{cm}(\uparrow\downarrow\rightarrow\downarrow s)\mathcal{\bar{M}}_{cm}(\downarrow\uparrow\rightarrow\downarrow s),\nonumber\\
	\Lambda^{1,3}_{\text{AC}}(s)&=&\mathcal{M}_{cm}(\uparrow\uparrow\rightarrow\uparrow s)[(c_{1}c_{3}+c_{5}c_{7})\mathcal{\bar{M}}_{cm}(\uparrow\downarrow\rightarrow\downarrow s)\nonumber\\
	&+&(c_{1}c_{5}+c_{3}c_{7})\mathcal{\bar{M}}_{cm}(\downarrow\uparrow\rightarrow\downarrow s)],\nonumber\\
	\Lambda^{1,4}_{\text{AC}}(s)&=&\mathcal{M}_{cm}(\uparrow\uparrow\rightarrow\uparrow s)[(c_{1}c_{4}+c_{5}c_{8})\mathcal{\bar{M}}_{cm}(\uparrow\downarrow\rightarrow\downarrow s)\nonumber\\
	&+&(c_{1}c_{6}+c_{3}c_{8})\mathcal{\bar{M}}_{cm}(\downarrow\uparrow\rightarrow\downarrow s)],\nonumber\\
	\Lambda^{2,3}_{\text{AC}}(s)&=&\mathcal{M}_{cm}(\uparrow\uparrow\rightarrow\uparrow s)[(c_{2}c_{3}+c_{6}c_{7})\mathcal{\bar{M}}_{cm}(\uparrow\downarrow\rightarrow\downarrow s)\nonumber\\
	&+&(c_{2}c_{5}+c_{4}c_{7})\mathcal{\bar{M}}_{cm}(\downarrow\uparrow\rightarrow\downarrow s)],\nonumber\\
	\Lambda^{2,4}_{\text{AC}}(s)&=&\mathcal{M}_{cm}(\uparrow\uparrow\rightarrow\uparrow s)[(c_{2}c_{4}+c_{6}c_{8})\mathcal{\bar{M}}_{cm}(\uparrow\downarrow\rightarrow\downarrow s)\nonumber\\
	&+&(c_{2}c_{6}+c_{4}c_{8})\mathcal{\bar{M}}_{cm}(\downarrow\uparrow\rightarrow\downarrow s)].\label{rhosA}
\end{eqnarray}

Then, tracing over $A$ using the orthogonality relation (\ref{traza}) yields the $2\times2$ reduced matrix for the system $C$,
\begin{align}
	\rho_{\text{C}}= \frac{1}{\mathcal{N}}\Big(\rho_{\text{C(in)}} + \rho_{\text{C(trans)}}\Big)\equiv\begin{pmatrix}
		\tilde{\Lambda}_C^{1,1} & \tilde{\Lambda}_C^{1,2}  \\\\
		\tilde{\Lambda}_C^{2,1} & \tilde{\Lambda}_C^{2,2}  \\
	\end{pmatrix},
	\label{rhoC}
\end{align}
with
\begin{equation}
	\rho_{\text{C(in)}}=8E_{\textbf{\text{q}}}E^{2}V^{3} \times\label{cin}
\end{equation}
\begin{equation*}\times
	\begin{pmatrix}
	c_{1}^{2}+c_{3}^{2}+c_{5}^{2}+c_{7}^{2} & c_{1}c_{2}+c_{3}c_{4}+c_{5}c_{6}+c_{7}c_{8} \\\\
	c_{1}c_{2}+c_{3}c_{4}+c_{5}c_{6}+c_{7}c_{8} & c_{2}^{2}+c_{4}^{2}+c_{6}^{2}+c_{8}^{2} 
	\end{pmatrix}
\end{equation*}
and 
\begin{equation}
	\rho_{\text{C(trans)}}=\frac{E_{\textbf{\text{q}}}TV^{2}|\textbf{\text{p}}_{3}|}{16\pi^{2}E}\SumInt_{s} d\Omega \times \label{trans}
\end{equation}
\vskip-0.5cm
\begin{equation*}\times	
	\begin{pmatrix}
		\Lambda^{1,1}_{\text{AC}}(s)+\Lambda^{3,3}_{\text{AC}}(s) & \Lambda^{1,2}_{\text{AC}}(s)+\Lambda^{3,4}_{\text{AC}}(s) \\\\ 
		\Lambda^{1,2}_{\text{AC}}(s)+\Lambda^{3,4}_{\text{AC}}(s) & \Lambda^{2,2}_{\text{AC}}(s)+\Lambda^{4,4}_{\text{AC}}(s) 
	\end{pmatrix}.
\end{equation*}

 In principle, if there are modifications in the matrix elements of the reduced density matrix $\rho_\text{C}$ after the scattering, as given by equations \eqref{rhoC}, \eqref{cin} and \eqref{trans}, the variation of the spin measurement on spectator particle $C$, reads
\begin{eqnarray}
&&\Delta\langle S_{x,y,z}\rangle = \langle S_{x,y,z}\rangle_{\mathrm{out}} - \langle S_{x,y,z}\rangle_{\mathrm{in}}\nonumber \\
&& \equiv\frac{1}{2} \text{Tr}[\sigma_{x,y,z} \,\, \rho_{\text{C}}] -\frac{1}{2} \text{Tr}[\sigma_{x,y,z} \,\, \rho_{\text{C(in)}}],
\end{eqnarray}
from which we may infer dynamical information about the scattering process, as we shall see in next sections. 
\section{Inelastic scattering $\boldsymbol{e^{+}e^{-}}\rightarrow\mu^{+}\mu^{-}$}\label{inelastic}
\begin{figure}[H]
\captionsetup{justification=centering}
\centering
\includegraphics[scale=0.43]{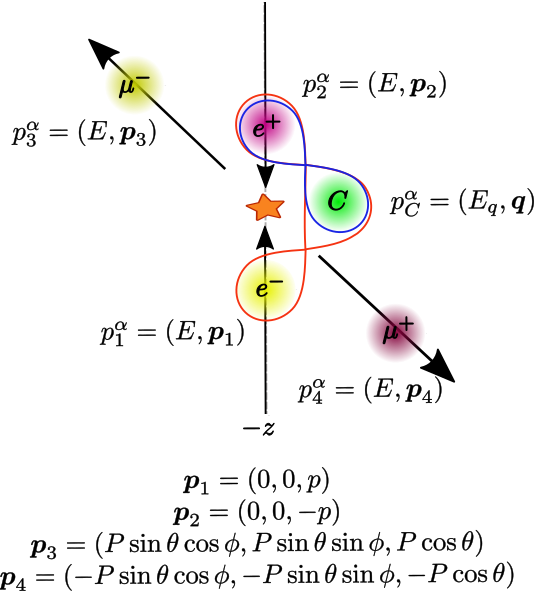} 
\caption{Collision diagram for the inelastic scattering $e^{+}e^{-}\rightarrow\mu^{+}\mu^{-}$.}
\label{collisiondiagram}
\end{figure}

\begin{figure}[H]
\captionsetup{justification=centering}
\centering
\includegraphics[scale=0.3]{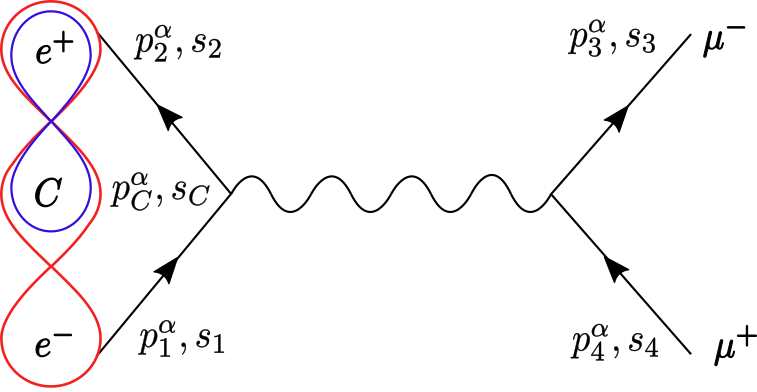} 
\caption{$s$ channel for the inelastic scattering $e^{+}e^{-}\rightarrow\mu^{+}\mu^{-}$.}
\label{fig:Feynman}
\end{figure}

Consider the inelastic scattering $e^{+}e^{-}\rightarrow\mu^{+}\mu^{-}$ in quantum electrodynamics taken in the centre-of-mass frame, where the electron/positron collision along the $z$-axis gives rise to a muon/anti-muon pair with centre-of-mass momenta $p$ and $P$, respectively as depicted in figure \ref{collisiondiagram}. The scattering angle is $\theta$, between $\bf{p_1}(\bf{p_2})$ and $\bf{p_3}(\bf{p_4})$ and $\phi$ is the azimuthal angle. In the initial state expressed by equation (\ref{eqn:estado-inicial}), we will admit a spectator particle $C$ (Claire)  which may be entangled in spin degrees of freedom with the positron (Bob) as represented by the blue curved line, or with both the positron and the electron (Alice) as represented by the red line, determined by specific values of $c_i$'s in (\ref{estadogeneralinicial}). For specific spin quantum numbers $s_i$, the tree level contribution to this scattering contains only the $s$-channel represented by the Feynman diagram depicted in figure \ref{fig:Feynman}, whose analytic expression reads
\begin{equation}
\mathcal{M}_{e^+e^-\rightarrow \mu^+\mu^-} = \frac{e^2}{(p_1^\alpha+p_2^\alpha)^2}[\bar{u}_3^{s_3}\gamma^\mu v_4^{s_4}][\bar{v}_2^{s_2}\gamma_\mu u_1^{s_1}],
\end{equation}
where
\begin{align*}
	u_{1}^{\uparrow}=N_{1}\begin{pmatrix}
		1\\
		0\\
		\frac{p}{E+m_{e}}\\
		0
	\end{pmatrix},\quad u_{1}^{\downarrow}=N_{1}\begin{pmatrix}
		0\\
		1\\
		0\\
		\frac{-p}{E+m_{e}}
	\end{pmatrix},
\end{align*}
\vskip-0.6cm
\begin{align*}
	\bar{v}_{2}^{\uparrow}=N_{1}\begin{pmatrix}
		0,\frac{p}{E+m_{e}},0,1
	\end{pmatrix}\gamma^{0},\quad \bar{v}_{2}^{\downarrow}=-N_{1}\begin{pmatrix}
		\frac{-p}{E+m_{e}},0,1,0
	\end{pmatrix}\gamma^{0},
\end{align*}
\begin{align*}
	\bar{u}_{3}^{\uparrow}=N_{2}\begin{pmatrix}
		1,0,\frac{P}{E+m_\mu}\cos\theta,\frac{P}{E+m_\mu}\sin\theta e^{-i\phi}
	\end{pmatrix}\gamma^{0},\nonumber\\ \bar{u}_{3}^{\downarrow}=N_{2}\begin{pmatrix}
		0 ,1 , \frac{P}{E+m_\mu}\sin\theta e^{i\phi} , \frac{-P}{E+m_\mu}\cos\theta
	\end{pmatrix}\gamma^{0},
\end{align*}
\begin{align}
	v_{4}^{\uparrow}=N_{2}\begin{pmatrix}
		\frac{-P}{E+m_\mu}\sin\theta e^{-i\phi}\\
		\frac{P}{E+m_\mu}\cos\theta\\
		0\\
		1
	\end{pmatrix}, v_{4}^{\downarrow}=-N_{2}\begin{pmatrix}
		\frac{-P}{E+m_\mu}\cos\theta\\
		\frac{-P}{E+m_\mu}\sin\theta e^{i\phi}\\
		1\\
		0
	\end{pmatrix},
\end{align}
with normalizations $N_{1}=\sqrt{E+m_{e}}$, and $N_{2}=\sqrt{E+m_\mu}$ in terms of the masses of the electron $m_{e}$ and the muon $m_\mu$. In order to the produce a $\mu^+\mu^-$ pair, the centre of mass energy $\sqrt{s}$ must be such that $s=(p_1+p_2)^2=4 E^2 \ge 4 m_\mu^2$.

\subsection{Entangled spin tripartite initial state}\label{totalentanglement}

\begin{figure}[H]
\centering
\captionsetup{justification=centering}
\includegraphics[width=8.7cm, height=6cm]{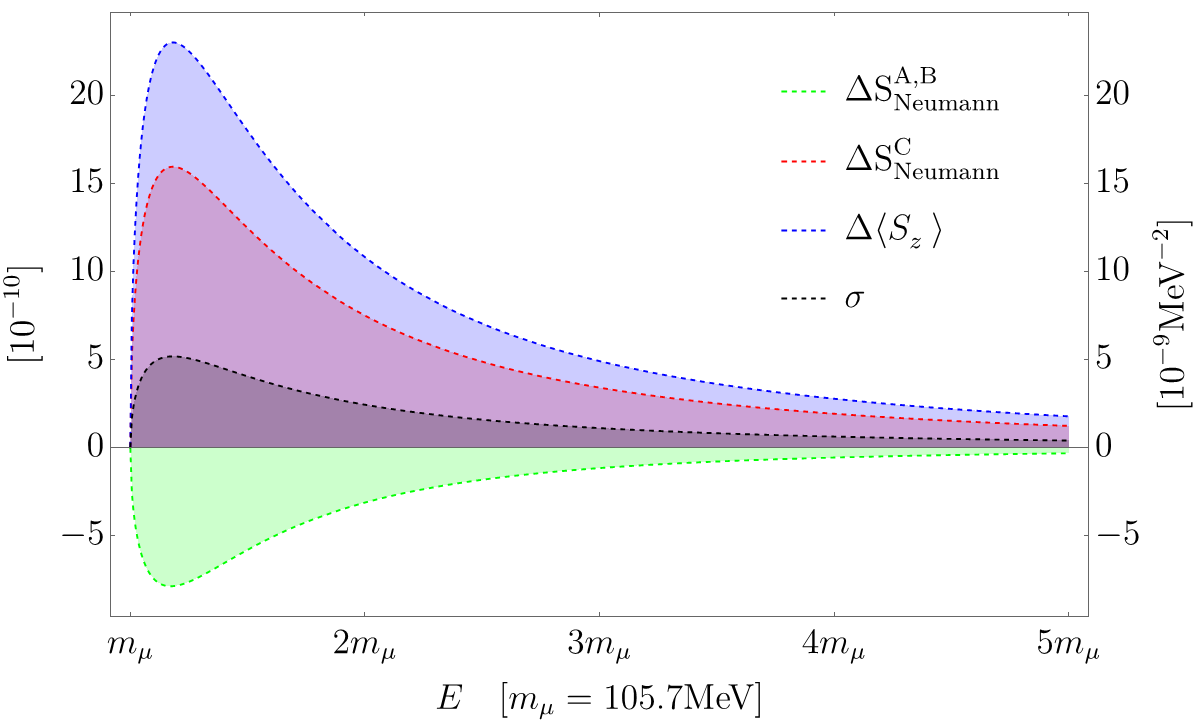}
\caption{Variations in the state W of the von Neumann entropy for $A,B$ and $C$ and the measured spin in the $z$-direction for $C$, as well as the cross section for the scattering $AB$.}
\label{fig:f2}
\end{figure}

\begin{figure}[H]
\centering
\captionsetup{justification=centering}
\includegraphics[width=8.5cm, height=6cm]{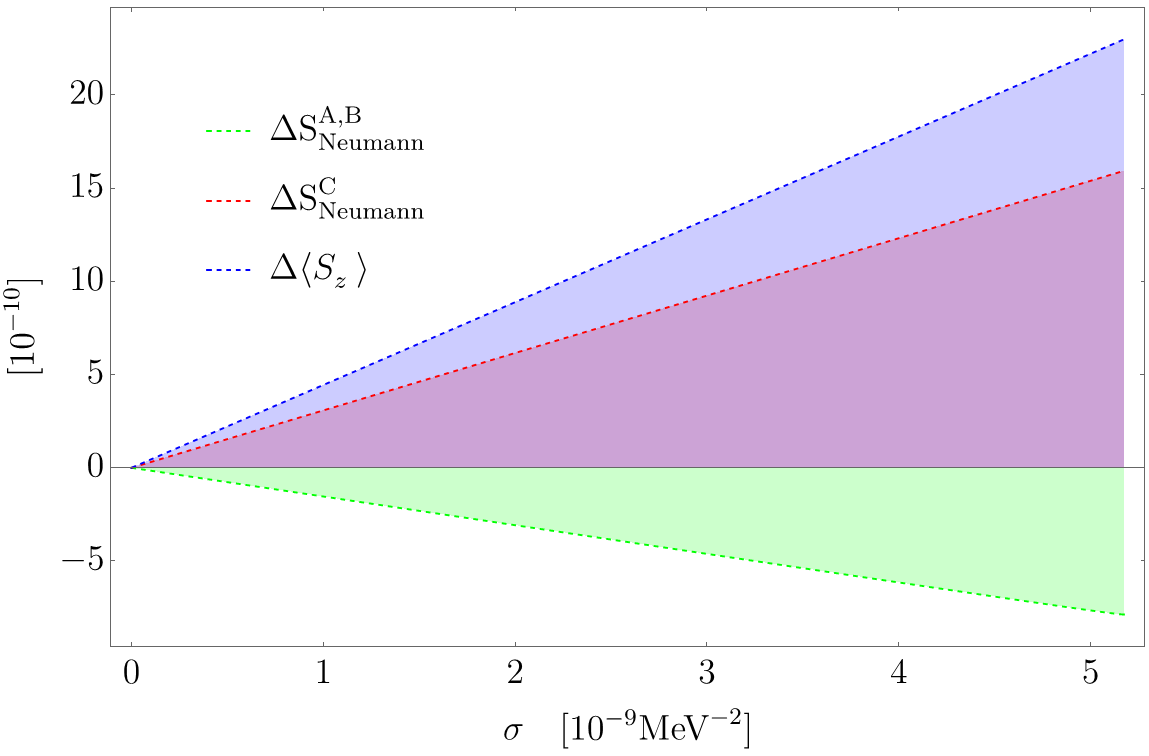}
\caption{Variations in the state W of the von Neumann entropies and the measured spin in the $z-$direction for $C$ in terms of the cross section.}
\label{fig:f3}
\end{figure}

Consider firstly GHZ and W initial spin in-states. In the scattering of $AB$ system, the composite states obey $\rho_{\text{AC}}^{\text{GHZ}}=\rho_{\text{BC}}^{\text{GHZ}}$ and $\rho_{\text{AC}}^{\text{W}}=\rho_{\text{BC}}^{\text{W}}$ before and after collision.
In other words, the collision affects equally the bi-partitions $AC$ and $BC$, resulting in equivalent spin density matrices. This result is expected because both $A$ and $B$ are equally entangled with $C$. On the other hand, the von Neumann entropy $\text{S}_{\text{Neumann}}$,
\begin{equation}
\text{S}^{\text{XC}}_{\text{Neumann}} = - \text{Tr} [\rho_{\text{XC}} \ln \rho_{\text{XC}}],
\end{equation}
$\text{X}=\{A,B\}$, will always increase (since the total system $ABC$ starts as a pure state with a null entropy \cite{horodick}), and it does so more significantly after the collision where there are more contributing states for all spin configurations defined by the constants $c_{(j)}$. This in general leads to a change in the entropy of individual systems. Let us specifically study the variation 
\begin{equation}
\Delta\text{S}^{\text{C}}_{\text{Neumann}} = (\text{S}^{\text{C}}_{\text{Neumann}})_{\text{out}} - (\text{S}^{\text{C}}_{\text{Neumann}})_{\text{in}}
\end{equation}
with $\text{S}^{\text{C}}_{\text{Neumann}} =-\text{Tr} [\rho_\text{C} \ln \rho_\text{C}],$ for a spin configuration such that the in-state is characterised by 
GHZ and W tripartite spin states. For the state GHZ, although the bipartite $AC$ or $BC$ state becomes more mixed, the entropy of the reduced system will not change as it started at its maximal mixedness degree. Hence the von Neumann entropy of the spectator particle $C$ will not be affected by the collision  for the GHZ configuration. On the other hand for the state W, the systems do not  start with a maximum mixedness, and thus the collision at centre-of-mass energy $E$ can increase the entropy of particle $C$. We can schematically represent these features in terms of the reduced density matrices. For the GHZ state,
\begin{align}
\begin{pmatrix}
\frac{1}{2} & 0 \\
0 & \frac{1}{2}
\end{pmatrix}_{\text{GHZ}}^{\text{A,B,C}}\xrightarrow{\text{Scattering}}\begin{pmatrix}
\frac{1}{2} & 0 \\
0 & \frac{1}{2}
\end{pmatrix}_{\text{GHZ}}^{\text{A,B,C}}\label{ghzall}
\end{align}
whereas for the W state, for the systems $A$ and $B$,
\begin{align}
\begin{pmatrix}
\frac{1}{3} & 0 \\
0 & \frac{2}{3}
\end{pmatrix}_{\text{W}}^{\text{A,B}}\xrightarrow{\text{Scattering}}
\begin{pmatrix}
f_1(E) & 0\\
0 & f_2(E)
\end{pmatrix}_{\text{W}}^{\text{A,B}}
\end{align}
in which $f_1 (E)$ and $f_2 (E)$ are functions of the energy $E$, and for the spectator particle C,
\begin{align}
\begin{pmatrix}
\frac{1}{3} & 0 \\
0 & \frac{2}{3}
\end{pmatrix}_{\text{W}}^{\text{C}}\xrightarrow{\text{Scattering}}
\begin{pmatrix}
f_3(E) & 0\\
0 & f_4(E)
\end{pmatrix}_{\text{W}}^{\text{C}}
\end{align}
where here we write out
\begin{equation}
f_3 (E) \equiv \frac{24\pi VE^{7}+E^{2}\mathscr{E}^{2}}{72\pi VE^{7}+(2m^{2}_{e}+E^{2})\mathscr{E}^{2}}\label{f3}
\end{equation}
and
\begin{equation}
f_4 (E) \equiv \frac{48\pi VE^{7}+2m_{e}^{2}\mathscr{E}^{2}}{72\pi VE^{7}+(2m^{2}_{e}+E^{2})\mathscr{E}^{2}},\label{f4}
\end{equation}
with $\mathscr{E}^{2}\equiv e^{4}T\sqrt{E^{2}-m_\mu^{2}}(m_\mu^{2}+2E^{2})$. The graph in figure \ref{fig:f2} shows the variations of the von Neumann entropy for the three systems and the measured spin of the spectator particle $C$ in the $z$-direction, as well as the cross section for the $AB$ collision, which in the center-of-mass frame reads:
\begin{eqnarray}
\sigma &=& \frac{1}{64\pi^2 s} \frac{P}{p} \int d\Omega \, |\mathcal{M}_{\text{i}\rightarrow\text{f}}|^{2}\label{crosssection}
\nonumber \\ &\equiv& \frac{e^{4}(m_{e}^{2}+E^{2})(m_{\mu}^{2}+2E^{2})}{48\pi E^{6}}\sqrt{\frac{E^{2}-m_{\mu}^{2}}{E^{2}-m_{e}^{2}}},\label{crosssectionW} 
\end{eqnarray}
all as a function of the energy $E$. We display in the left vertical axis the values associated with the dimensionless quantities $\Delta \text{S}_{\text{Neumann}}$ and $\Delta \langle S_{z} \rangle$ ($\hbar=1$), whilst the right vertical axis corresponds to the total cross section values in  $\text{MeV}^{-2}$. Notice that as no off-diagonal elements are developed for the reduced matrix $\rho_\text{C}$, $\Delta \langle S_{x,y} \rangle = 0$. Furthermore, starting from the threshold energy for this process where $E = m_\mu$, figure \ref{fig:f2} shows a common peak around $1.18 m_\mu$ for the four curves.
It is also possible to relate the variations of the von Neumann entropies of the three systems and the variation of the spin expectation value of $C$ in the $z$-direction with the cross section, the last two quantities as observables, since they can be expressed in terms of the energy $E$ via equations \eqref{f3}, \eqref{f4} and \eqref{crosssectionW}. This is displayed in figure \ref{fig:f3}.
Moreover, in Ref. \cite{FAN1}, the inelastic scattering $e^+ e^- \rightarrow \mu^+ \mu^-$ was studied and
the variation of entanglement entropy between the initial state and final state was seen to be proportional
 to the total cross section. In our case, we can also see in figure \ref{fig:f3} that the variation of von Neumann entropy of the three systems, including the  spectator particle, are proportional to the cross section.

\subsection{Bob and Claire initial entangled spin state}\label{partialentanglement}

\begin{figure}[H]
\centering
\captionsetup{justification=centering}
\includegraphics[width=8.5cm, height=6.5cm]{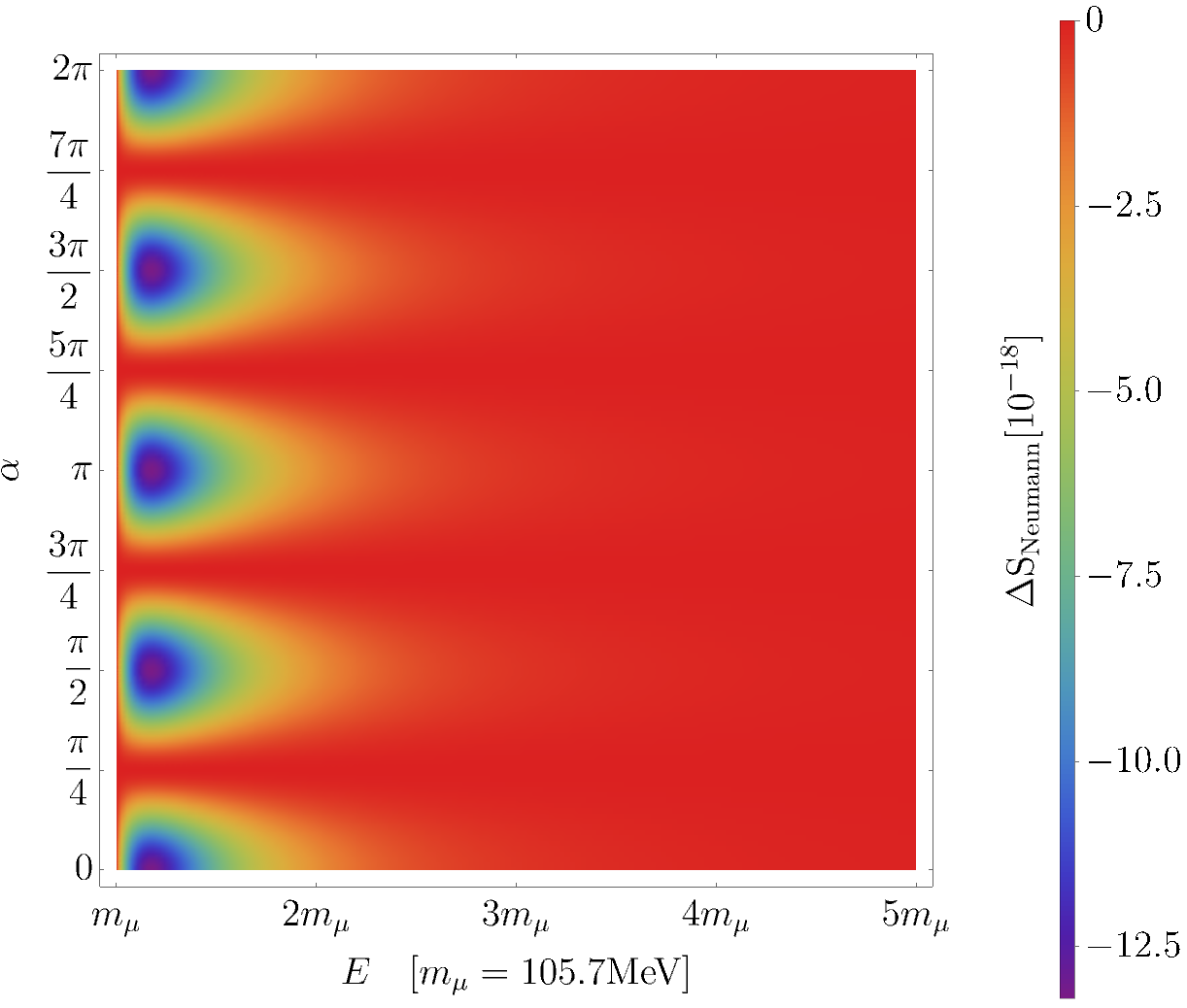}
\caption{Variation of the von Neumann entropy for $C$ in the state $\ket{\textup{A}^{\alpha}}\otimes\ket{\Psi^{\pm}(\Phi^{\pm})}$.}
\label{fig:f4}
\end{figure}

Contrarily to the GHZ and W initial entangled states, in this case the bipartitions $AC$ and $BC$ are not symmetrical as $C$ is initially entangled with $B$. That is because $AC$ begins as a mixed state namely $\rho_{\text{AC}}=\rho_{\text{A}}\otimes\rho_{\text{C}}$, where $\rho_{\text{A}}$ is a pure and $\rho_{\text{C}}$ (just as $\rho_{\text{B}}$) is a mixed state derived of entanglement, whereas BC begins as a pure entangled state. The degree of the mixedness associated to the spin density matrix of any individual system ($A$, $B$ and $C$) must comply with the fact that the mixedness of bipartitions ($AC$, $BC$ and $AB$) increase due to the scattering. In the case of the system $C$, the mixedness as measured by the von Neumann entropy clearly decreases as seen in  figure \ref{fig:f4}. 
For $\alpha=n\pi/4$, $n$ being and integer number, this variation decreases with the energy $E$ when $n$ is even and vanishes for $n$ odd. In our general state $\ket{\text{A}^{\alpha}}\otimes\ket{\Psi^{\pm}(\Phi^{\pm})}$ state,  system $A$ starts off maximally pure, that is its initial entropy $\text{S}^\text{A}_{\text{Neumann}}$ is zero. 
On the other hand given that the mixedness of the bipartitions $AB$ and $AC$ must increase after the scattering, so does the entropy of system $A$ (as it cannot decrease).  

On the other hand, for the individual systems $B$ and $C$ the opposite happens: they cannot increase their mixedness beyond the limit $\text{S}_{\text{Neumann}}^{\text{B,C}}=\ln2$, and thus a decrease of their mixedness is the only way to satisfy the constraints imposed by the increase of mixedness of the respective composite bipartite systems due to scattering. 

\begin{figure}[H]
\captionsetup{justification=centering}
\begin{subfigure}[H]{0.48\textwidth}
\includegraphics[width=8.5cm, height=6.5cm]{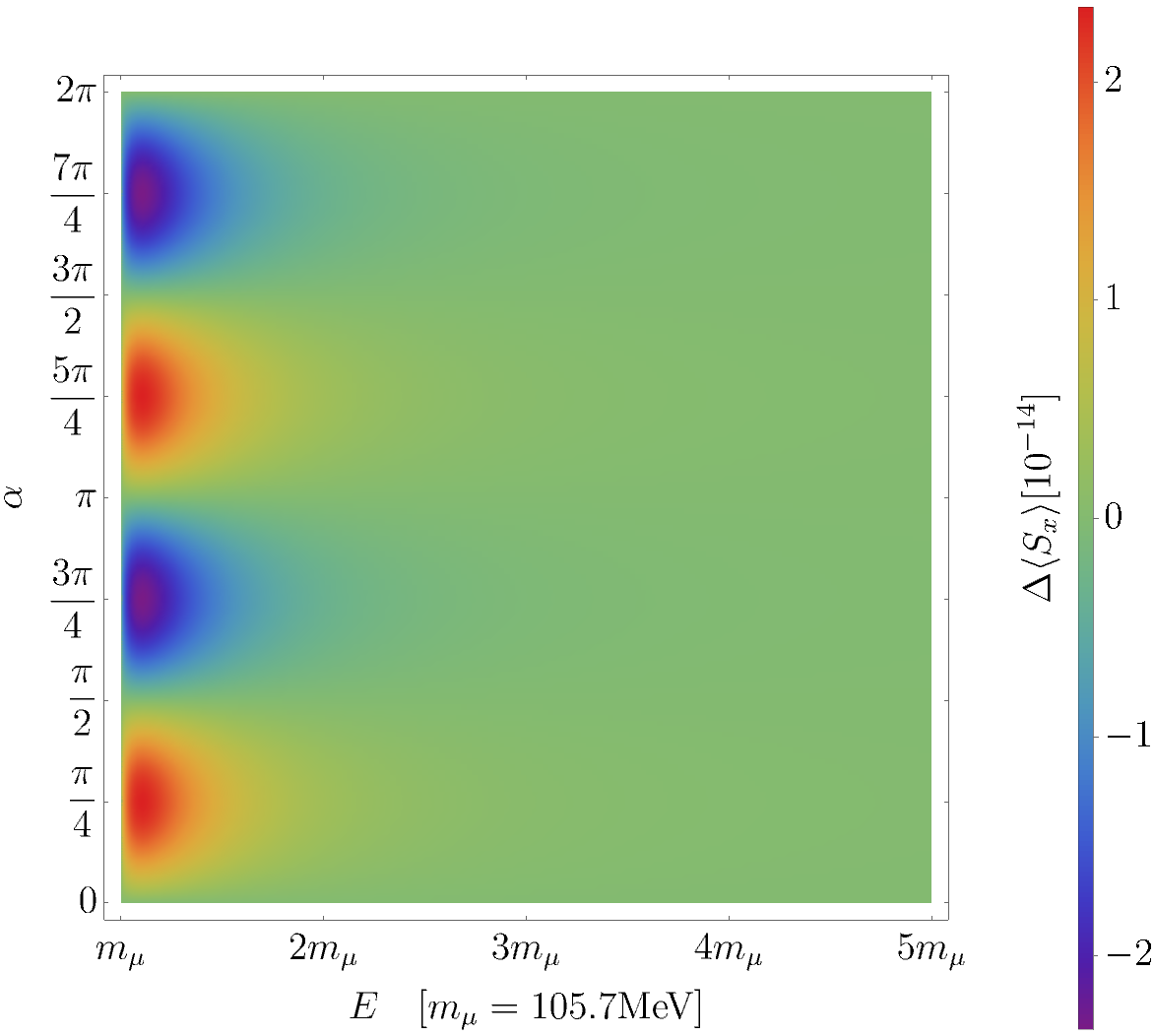}
\caption{$\ket{\textup{A}^{\alpha}}\otimes\ket{\Psi^{+}(\Phi^{+})}$ state.}
\label{fig:f5}
\end{subfigure}
\hfill
\begin{subfigure}[H]{0.48\textwidth}
\includegraphics[width=8.5cm, height=6.5cm]{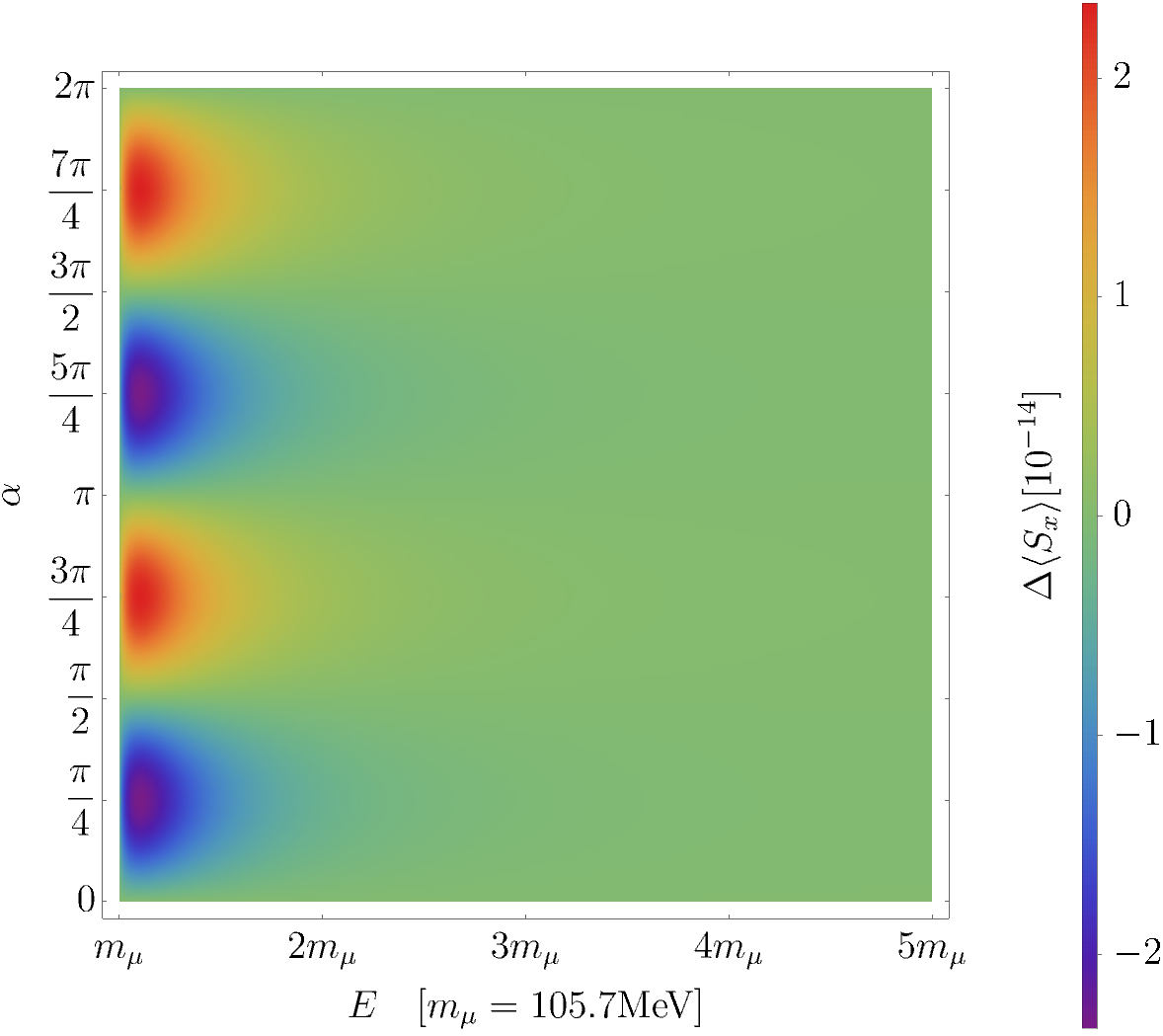}
\caption{$\ket{\textup{A}^{\alpha}}\otimes\ket{\Psi^{-}(\Phi^{-})}$ state.}
\label{fig:f6}
\end{subfigure}
\caption{Variation of the measured spin in the $x-$direction for $C$ in the states $\ket{\textup{A}^{\alpha}}\otimes\ket{\Psi^{\pm}(\Phi^{\pm})}$.}
\label{fig:f6T}
\end{figure}

The effect of the parameter $\alpha$ on the variation of spin expectation value for system $C$ is deduced from its reduced density matrices:
\begin{align}
\begin{pmatrix}
\frac{1}{2} & 0 \\
0 & \frac{1}{2}
\end{pmatrix}_{\Psi^{\pm}}^{\text{C}}\xrightarrow{\text{Scattering}}
\begin{pmatrix}
g_{1}(E,\alpha) & \frac{1}{2}\sin(2\alpha)g_{3}^{\pm}(E)  \\\\
\frac{1}{2}\sin(2\alpha)g_{3}^{\pm}(E)  & g_{2}(E,\alpha) 
\end{pmatrix}_{\Psi^{\pm}}^{\text{C}}\label{PSIc}
\end{align}
\begin{align}
\begin{pmatrix}
\frac{1}{2} & 0 \\
0 & \frac{1}{2}
\end{pmatrix}_{\Phi^{\pm}}^{\text{C}}\xrightarrow{\text{Scattering}}
\begin{pmatrix}
g_{2}(E,\alpha) & \frac{1}{2}\sin(2\alpha)g_{3}^{\pm}(E) \\\\
\frac{1}{2}\sin(2\alpha)g_{3}^{\pm}(E) & g_{1}(E,\alpha)
\end{pmatrix}_{\Phi^{\pm}}^{\text{C}}\label{PHIc}
\end{align}
where
\begin{equation}
g_{1}(E,\alpha) =\frac{48\pi VE^{7}+\mathscr{E}^{2}[m^{2}_{e}\cos^{2}\alpha+2E^{2}\sin^{2}\alpha]}{96\pi VE^{7}+(m^{2}_{e}+2E^{2})\mathscr{E}^{2}},\label{g1}
\end{equation}
\begin{equation}
g_{2}(E,\alpha) =\frac{48\pi VE^{7}+\mathscr{E}^{2}[m^{2}_{e}\sin^{2}\alpha+2E^{2}\cos^{2}\alpha]}{96\pi VE^{7}+(m^{2}_{e}+2E^{2})\mathscr{E}^{2}},\label{g2}
\end{equation}
\begin{equation}
g_{3}^{\pm}(E) =\pm\frac{m^{2}_{e}\mathscr{E}^{2}}{96\pi VE^{7}+(m^{2}_{e}+2E^{2})\mathscr{E}^{2}};\label{g3}
\end{equation}
the superscript $+$ ($-$) refers to $\Psi^+,\Phi^+$ ($\Psi^-,\Phi^-$). Interestingly, the off-diagonal values of the reduced density matrix for the system $C$ in equations \eqref{PSIc} and \eqref{PHIc} are identical and proportional to the off-diagonal elements of the density matrix for system $A$ before scattering:
\begin{align*}
\begin{pmatrix}
\cos^{2}\alpha & \frac{1}{2}\sin(2\alpha) \\\\
\frac{1}{2}\sin(2\alpha) & \sin^{2}\alpha
\end{pmatrix}_{\Psi^{\pm},\Phi^{\pm}}^{\text{A}}
\end{align*}
In other words, the $\alpha$ parameter which specifies the spin state of the system $A$ has an important role in the computation of spin expectation values for the system $C$ in the $x$-direction after the scattering. Therefore we expect the same behavior of $\Delta \langle S_x \rangle $ for $\ket{\text{A}^\alpha}\otimes\ket{\Psi^{+}}, \ket{\text{A}^\alpha}\otimes\ket{\Phi^{+}}$ as well as for $\ket{\text{A}^\alpha}\otimes\ket{\Psi^{-}}, \ket{\text{A}^\alpha}\otimes\ket{\Phi^{-}}$ as can be seen in figures \ref{fig:f5} and \ref{fig:f6}. When the systems $B$ and $C$ are coupled in the Bell bases $\ket{\Psi^{+}(\Phi^+)}$ the dependence on  $\alpha$ is such that for $\alpha=n\pi/4$, $n$ being and integer number, $\Delta \langle S_x \rangle$  decreases with the energy $E$ when $n$ is odd (with maxima and minima close to $E = m_\mu$) and vanishes for $n$ even. For the Bell bases $\ket{\Psi^{-}(\Phi^-)}$ we have a similar pattern, except for the interchange between maxima and minima with respect to the other Bell basis.

\begin{figure}[H]
\captionsetup{justification=centering}
\begin{subfigure}[b]{0.48\textwidth}
\includegraphics[width=8.5cm, height=6.5cm]{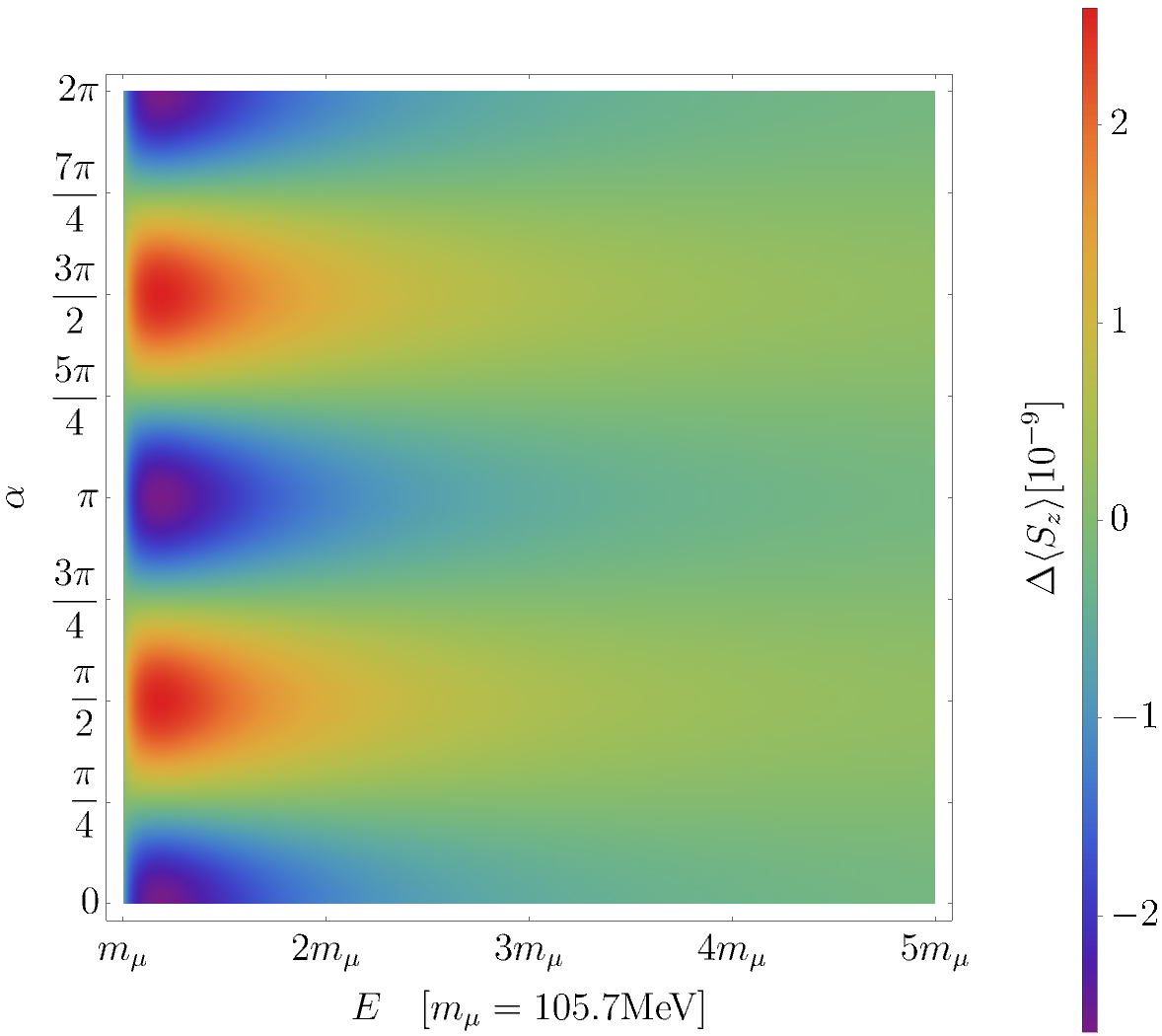}
\caption{$\ket{\textup{A}^{\alpha}}\otimes\ket{\Psi^{\pm}}$ state.}
\label{fig:f7}
\end{subfigure}
\hfill
\begin{subfigure}[b]{0.48\textwidth}
\includegraphics[width=8.5cm, height=6.5cm]{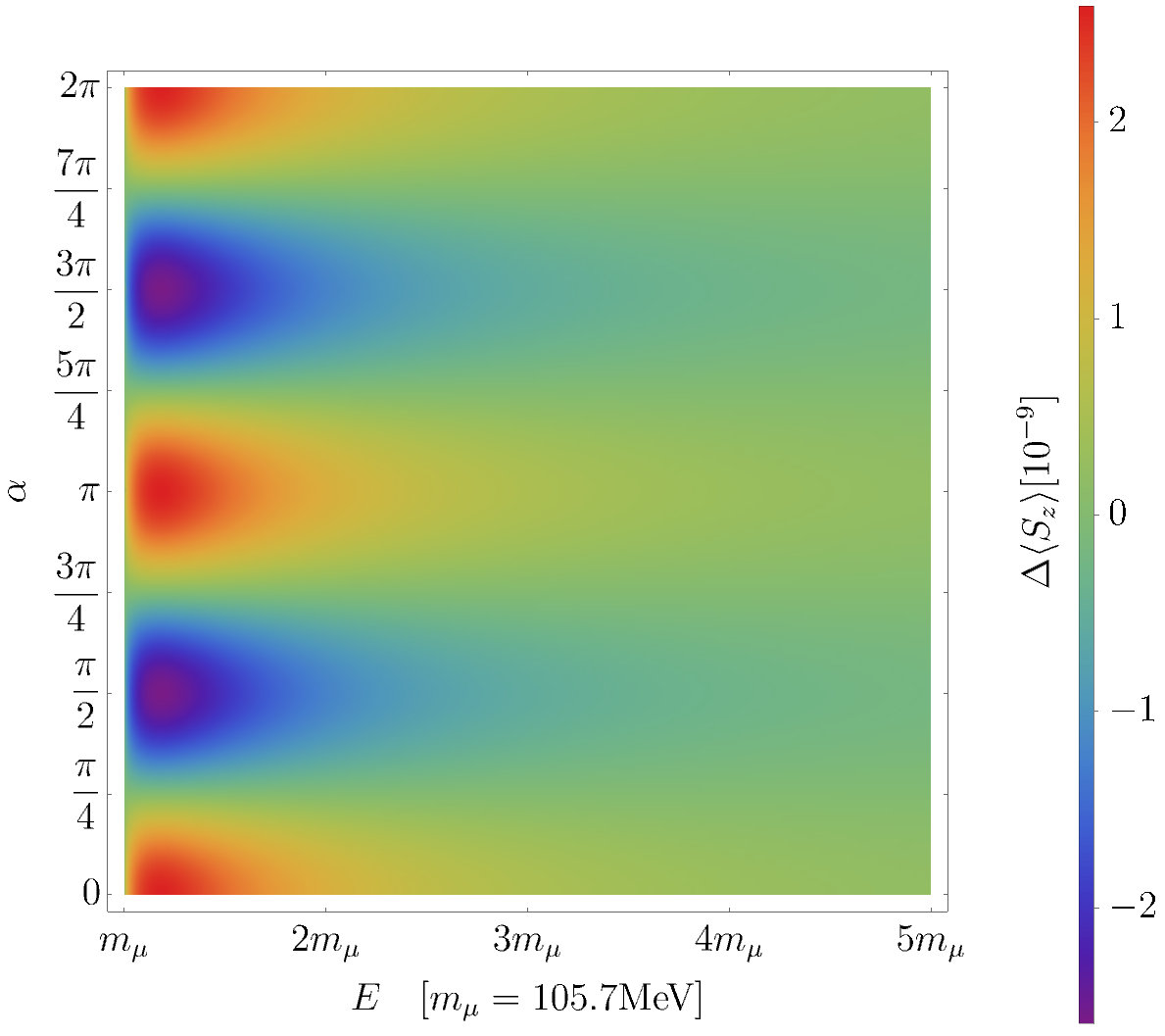}
\caption{$\ket{\textup{A}^{\alpha}}\otimes\ket{\Phi^{\pm}}$ state.}
\label{fig:f8}
\end{subfigure}
\caption{Variation of the measured spin in the $z-$direction for $C$ in the states $\ket{\textup{A}^{\alpha}}\otimes\ket{\Psi^{\pm}(\Phi^{\pm})}$.}
\label{fig:f8T}
\end{figure}

Recall that for the initial state discussed in the previous section, namely the state W, the reduced density matrix of system $C$ has not developed off-diagonal elements after the scattering and hence $\Delta \langle S_x \rangle = 0$. Notice that diagonal elements of the reduced density matrix $C$ in equations \eqref{PSIc} and \eqref{PHIc} are defined by the functions $g_{1,2} (E, \alpha)$ in \eqref{g1} and \eqref{g2}. Such diagonal elements, which play a key role in the determination of $\Delta\langle S_z^\text{C} \rangle$, have terms proportional to diagonal elements of the initial density matrix of the system $A$, namely $\cos^2\alpha$ and $\sin^2 \alpha$ besides a term proportional to $E^7$. 
These diagonal elements are interchanged in the states $\ket{\Psi^{\pm}}$ and $\ket{\Phi^{\pm}}$ as can be seen from the density matrices given by the equations \eqref{PSIc} and \eqref{PHIc}, respectively.  
This feature can be observed in the graphs depicted in figure \ref{fig:f7} and \ref{fig:f8}, where the variation of the spin measured for system $C$ in the $z$-direction is plotted as a function of the energy $E$ and the angle $\alpha$ that characterises the initial spin state of system $A$. For the Bell bases $\ket{\Psi^\pm}$ taking for $\alpha=n\pi/4$, $\Delta \langle S_z \rangle$  decreases with the energy $E$ when $n$ is even (with maxima and minima ranging from $E = m_\mu$ to $E = 1.5 m_\mu$) and vanishes for $n$ odd. For the Bell bases $\ket{\Phi^\pm}$ in figure \ref{fig:f8} we have a similar pattern, except for the interchange between maxima and minima with respect to figure \ref{fig:f7}. It is noteworthy that the $\Delta \langle S_z \rangle$ is 5 orders of magnitude larger than $\Delta \langle S_x \rangle$ in the Bell state configurations and 1 order of magnitude larger than $\Delta \langle S_z \rangle$ calculated for particle $C$ taking a state W as the initial spin state. 

We may summarize the case in which Bob and Claire are initially entangled whereas Alice is in a superposition spin state in the following table 

\onecolumngrid

\begin{table}[H]
\centering
\captionsetup{justification=centering}
\begin{TAB}(r,1cm,0.5cm)[2pt]{|c|c|c|c|c|c|}{|c|c|}
\textbf{Initial Spin of A} & $\boldsymbol{\braket{S_{x}^{\text{A}}}}$ & $\boldsymbol{\Delta\braket{S_{x}^{\text{C}}}}$ in $\boldsymbol{\ket{\Psi^{\pm}(\Phi^{\pm})}}$ & $\boldsymbol{\braket{S_{z}^{\text{A}}}}$ & $\boldsymbol{\Delta\braket{S_{z}^{\text{C}}}}$ in $\boldsymbol{\ket{\Psi^{\pm}}}$ & $\boldsymbol{\Delta\braket{S_{z}^{\text{C}}}}$ in $\boldsymbol{\ket{\Phi^{\pm}}}$ \\
$\cos\alpha\ket{\uparrow}+\sin\alpha\ket{\downarrow}$ & $\frac{1}{2}\sin(2\alpha)$ & $\frac{1}{2}\sin(2\alpha)\cdot g_{3}^{\pm}(E)$ & $\frac{1}{2}\cos(2\alpha)$ & $\frac{1}{2}\cos(2\alpha)\cdot h_{3}(E)$ & $-\frac{1}{2}\cos(2\alpha)\cdot h_{3}(E)$ \\
\end{TAB} 
\caption{$\alpha$-dependence in the spin expectation values of the spectator particle $C$ given the initial state of the system $A$ governed by $\alpha$.}
\label{table1}
\end{table}
\twocolumngrid
where
\begin{equation}
h_{3}(E)=\frac{(m^{2}-2E^{2})\mathscr{E}^{2}}{96\pi VE^{7}+(m^{2}+2E^{2})\mathscr{E}^{2}}. 
\end{equation}

\begin{figure}[H]
\centering
\captionsetup{justification=centering}
\includegraphics[width=8.7cm, height=6cm]{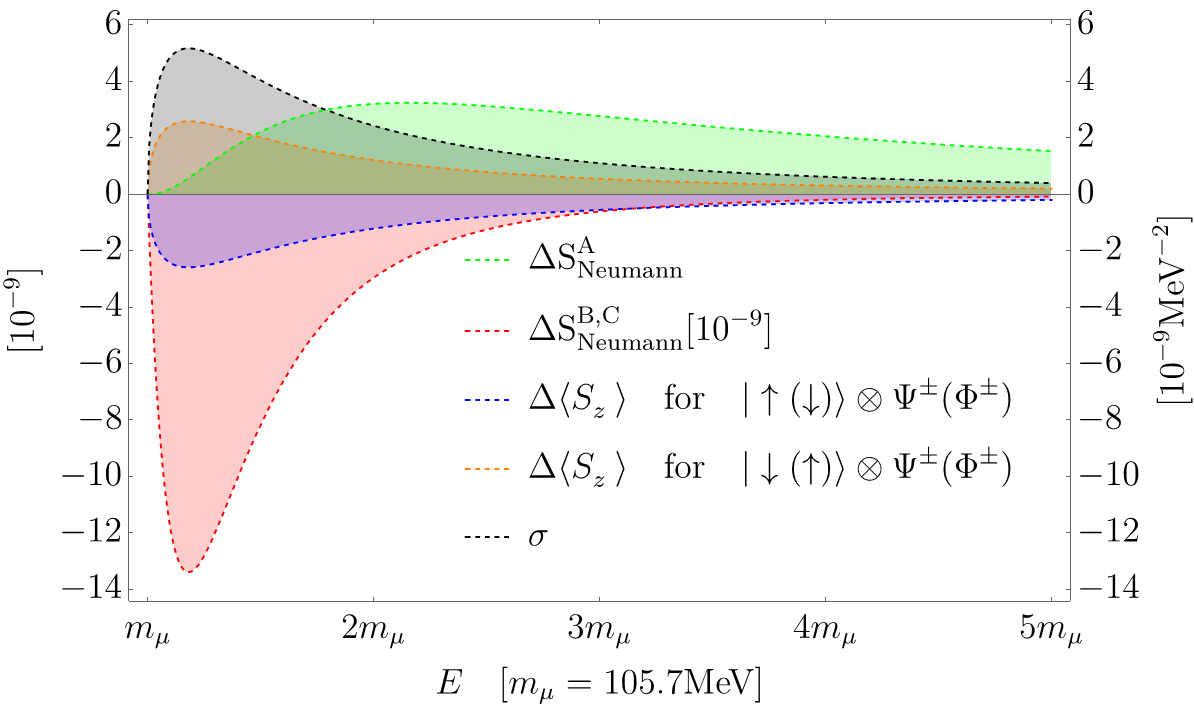}
\caption{Variations in the states $\ket{\textup{A}^{\alpha}}\otimes\ket{\Psi^{\pm}(\Phi^{\pm})}$ of the von Neumann entropy for $A,B$ and $C$ and the measured spin in the $z$-direction for $C$, as well as the cross section for the scattering $AB$.}
\label{fig:f9}
\end{figure}

Just as we have done for the W state, it is possible to parametrize the cross section together with the variations of von Neumann entropies and the expectation value of the spectator particle $C$ through the energy $E$ and $\alpha$. However, $\alpha$ is chosen such that $A$ is initially polarized given the prevalence of the values in $z-$direction with respect to $x-$direction as seen in the figures \ref{fig:f6T} and \ref{fig:f8T}. For this, we set $\alpha=0,\pi/2$ and display all these quantities in terms of the energy $E$ as shown in figure \ref{fig:f9}. As before, we display in the left vertical axis the values associated with the dimensionless quantities $\Delta \text{S}_{\text{Neumann}}$ and $\Delta \langle S_{z} \rangle$ ($\hbar=1$), whilst the right vertical axis corresponds to the total cross section values in  $\text{MeV}^{-2}$.

\begin{figure}[H]
\centering
\captionsetup{justification=centering}
\includegraphics[width=8.7cm, height=6cm]{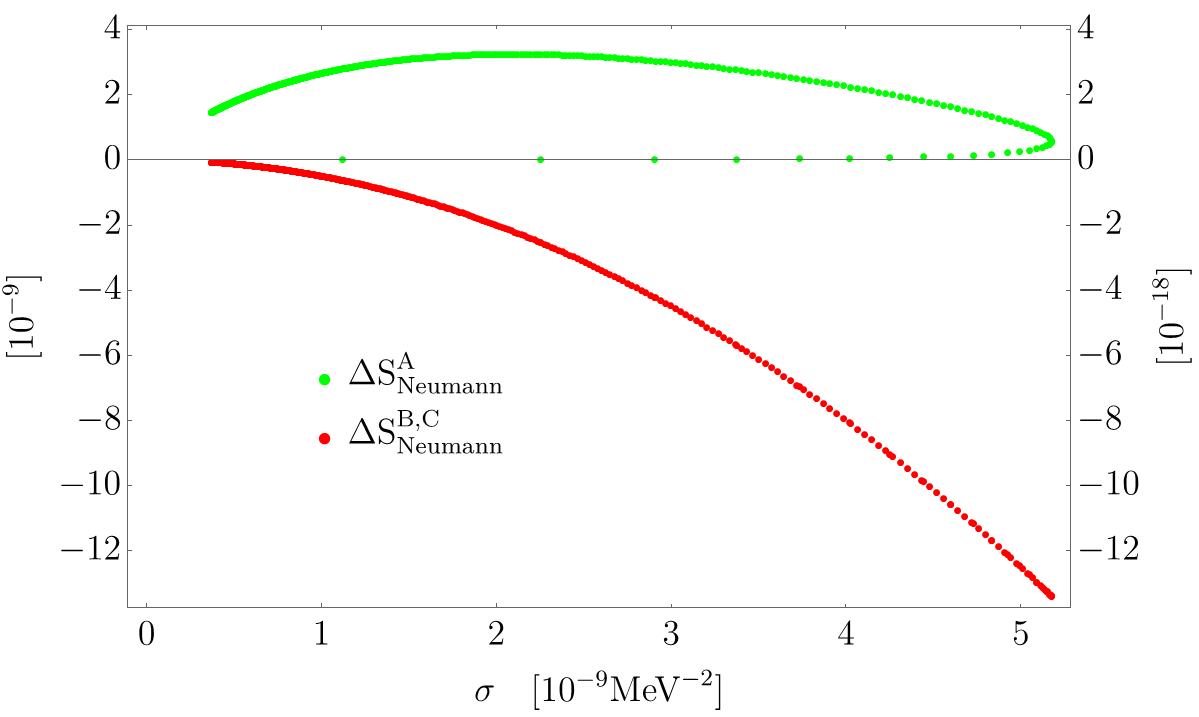}
\caption{$\Delta\text{S}_{\text{Neumann}}$ for the three systems as a function of the cross section for the states $\ket{\Psi^{\pm}(\Phi^{\pm})}$. We display in the left vertical axis the values associated with the variation of the von Neumann entropy for the system $A$, whilst the right vertical axis corresponds to systems $B$ and $C$.}
\label{fig:f10}
\end{figure}

\begin{figure}[H]
\centering
\captionsetup{justification=centering}
\includegraphics[width=8.2cm, height=6cm]{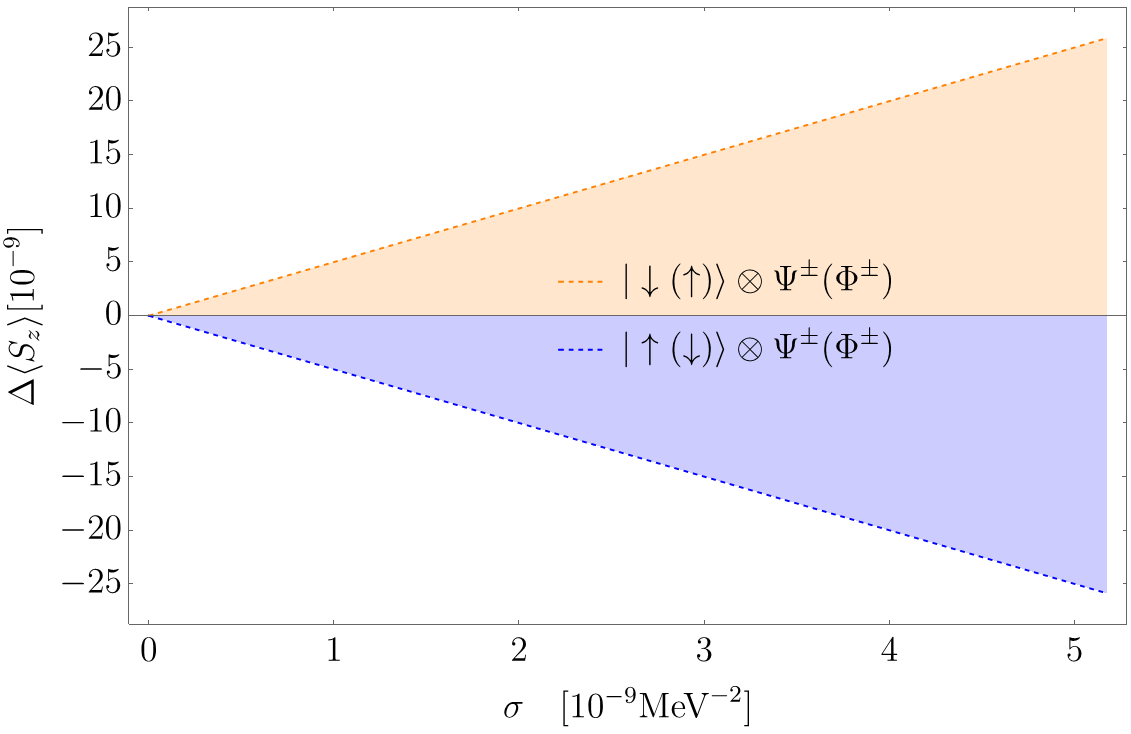}
\caption{$\Delta\braket{S_z}$ for the spectator particle $C$ as a function of the cross section for the states $\ket{\Psi^{\pm}(\Phi^{\pm})}$.}
\label{fig:f11}
\end{figure}

Notice that the entropy variations are not altered by choosing $\alpha=0$ or $\alpha=\pi/2$ as seen in the figure \ref{fig:f4} for $C$. This does not happen for the expectation spin value variations, where clearly there is a distinction for $\alpha=0$ and $\alpha=\pi/2$, also evidenced in the figures \ref{fig:f7} and \ref{fig:f8}. Furthermore, note that the peak of the entropy variation for $A$ corresponds to a different energy from that corresponding to the cross section; this makes the relation $\Delta\text{S}_{\text{Neumann}}^{\text{A}}$ vs $\sigma$ not univocal, while $\Delta\text{S}_{\text{Neumann}}^{\text{B,C}}$ vs $\sigma$ in fact is. This is shown in figure \ref{fig:f10}. The decrease curve in the variation of entropy for $B,C$ is just due to the difference between the orders of magnitude; i.e., $10^{-18}$ in $\Delta\text{S}_{\text{Neumann}}^{\text{B,C}}$ and $10^{-9}$ in $\sigma$. On the other hand, the relation $\Delta\braket{S_z}$ vs $\sigma$ is shown in figure \ref{fig:f11}, where clearly there is a linear proportionality and an exchange for $\alpha=0$ and $\alpha=\pi/2$. These are due to the coincidence between the peaks of both variables and the orders of magnitude.

\section{Discussion and perspectives}\label{conclusions}

To summarize, we have considered a general entangled state in spin degrees of freedom in a three particle system as the initial state of a QED S-matrix  calculation in which only two of the particles underwent dynamical scattering. Having constructed the corresponding reduced density matrices, we proceeded with the cross section calculation in the center-of-mass frame by specifying a QED inelastic tree-level process $e^+e^-\rightarrow \mu^+\mu^-$. We studied two specific classes of tripartite spin entanglement, namely GHZ and W states, and a class of states in which the spectator particle $C$ is entangled with one of the scattered particles (say,  particle $B$) in a Bell state. In the latter case, particle $A$ is assumed to be in a linear superposition of spin states (up and down) parameterized  by the angle $\alpha$.

For these configurations, the variation of the von Neumann entropies $\Delta \text{S}_{\text{Neumann}}^{\text{A,B,C}}$ and the variation of the expectation value of the $i^{th}$ component of spin $\Delta\langle S_i \rangle_C$  were evaluated. For specific configurations, such quantities encode information about the scattering dynamics via spin entanglement of the initial state. A natural question was whether or not  the reconstruction of a scattering physical observable, namely the total cross section, could be achieved by spin measurements on $C$.

 For the GHZ state, the mixedness of particle $C$ state measured by the von Neumann entropy, $\Delta\text{S}_{\text{Neumann}}$, does not change as the reduced spin density matrix remains unaltered after scattering. In this case, particle $C$ has no information about the collision between $A$ and $B$.
 
 However,  for the W state, the mixedness of  particle $C$ is increased through the emergence of energy dependent ($E$) diagonal elements in its spin density matrix which led to  a variation in its spin expectation value that also depends on $E$.  It is noteworthy that, as a function of the $E=\sqrt{s}/2$, the total cross section starts at  $E=m_\mu$, and grows until a peak around $E=1.18 \, m_\mu$ where it decreases monotonically to zero as $E \rightarrow \infty$. Taking the electron-positron pair initially in the $z$-direction,  $\Delta\text{S}_{\text{Neumann}}^{\text{C}}$ and $\Delta\langle S_{z}^{\text{C}}\rangle$, also starts at $E=m_\mu$, peaks at $E=1.18 \, m_\mu$, and decreases monotonically to zero as $E \rightarrow \infty$. On the other hand, the variations of the von Neumann entropies for particles $A$ and $B$, $\Delta\text{S}_{\text{Neumann}}^{\text{A,B}}$ had a minimum at $E=1.18 \, m_\mu$, and increased monotonically to zero as $E \rightarrow \infty$ as seen in Fig. \ref{fig:f2}. This paired behavior made it possible to obtain a direct relation between $\Delta\text{S}_{\text{Neumann}}^{\text{A,B,C}}$ and $\Delta\langle S_{z}\rangle$ with the cross section $\sigma$ as seen Fig. \ref{fig:f3},  enabling us to obtain direct information about the scattering process from spin measurements on the spectator particle $C$. In other words,  for a W initial state, $\Delta \text{S}_{\text{Neumann}}^{\text{A,B,C}}$ and $\Delta \langle S_z\rangle$ are seen to be directly proportional to the total cross section $\sigma$.

As for the states $\ket{\text{A}^{\alpha}}\otimes\ket{\Psi^\pm}$ and $\ket{\text{A}^{\alpha}}\otimes\ket{\Phi^\pm}$, the mixedness of each particle changed through the emergence of diagonal and non-diagonal elements in their spin density matrix, which depends on $E$ and  $\alpha$. For particle $A$, since it started at $\text{S}_{\text{Neumann}}^{\text{A}}=0$ (minimal), its mixedness increases after scattering. In contrast, for particles $B$ and $C$ whose initial mixedness started at $\text{S}_{\text{Neumann}}^{\text{B,C}}=\ln2$ (maximal), it decreases. This was significantly evident for $\alpha=n\pi/2$, $n \in \mathbb{Z}$ , and for an energy $E=1.18m_{\mu}$ (where $\Delta\text{S}_{\text{Neumann}}^{\text{B,C}}$ is minimal). For particle $C$, this meant a significant variation of spin expectation value in the $z-$direction over the $x-$direction as shown in Figs. \ref{fig:f6T} and \ref{fig:f8T}, where there was a difference in 5 orders of magnitude. 
As a function of $E$  these variations in both directions started at $E=m_{\mu}$ and dropped to zero as $E\rightarrow\infty$. There also occurred positive/negative variations of spin expectation values in the $x-$direction for $\alpha=(2n+1)\pi/4$, whereas such variations manifested for $\alpha=n\pi/2$   in the $z-$direction. It is remarkable that the value of $\alpha$ affects the variations of the spin expectation values of particle $C$ after scattering, as displayed in  table \ref{table1}. On the other hand,  $\Delta\langle S_x\rangle$ (with a peak at $E=\pm1.10m_\mu$) grows and decays faster than $\Delta\langle S_z\rangle$ (with a peak at $E=\pm1.18m_\mu$) as seen in graphs of Figs. \ref{fig:f6T} and \ref{fig:f8T}.   

Because $\Delta\langle S_z \rangle$ is 5 orders of magnitude greater than $\Delta\langle S_x \rangle$,
we fixed $\alpha=0$ and $\alpha=\pi/2$, where the variation was maximal in order to have $\Delta\langle S_z \rangle$ as a function of $E$ only. Such values of $\alpha$ correspond to the upper ($\alpha = 0$) and lower ($\alpha = \pi/2$) graphs depicted in Fig. \ref{fig:f11} showing the proportionality between $\Delta\langle S_z \rangle$ and $\sigma$.

In Fig. \ref{fig:f9} we showed the behaviour of  $\Delta\text{S}_{\text{Neumann}}^{\text{A,B,C}}$, $\Delta\langle S_z\rangle$, and the cross section $\sigma$ as a function of the energy $E$.  All curves started at $E=m_\mu$ and approached zero as $E\rightarrow\infty$. In the case of particle $A$, the variation $\Delta\text{S}_{\text{Neumann}}^{\text{A}}$ had a maximum for $E=2.11m_\mu$ and decreased monotonically to zero as $E\rightarrow\infty$. For particles $B$ and $C$, as previously mentioned, $\Delta\text{S}_{\text{Neumann}}^{\text{B,C}}$ had a minimum for $E=1.18m_\mu$ and increased monotonically to zero as $E\rightarrow\infty$. 
The fact that $\sigma$ and $\Delta\text{S}_{\text{Neumann}}^{\text{B,C}}$ reached their maximum and minimum, respectively, for the same energy $E=1.18\, m_\mu$, makes the relation univocal just as for the W state. This was shown in figure \ref{fig:f10}). The curved shape was only due to the difference between orders of magnitude; i.e., $10^{-9}[\text{MeV}^{-2}]$ for $\sigma$ and $10^{-18}$ for $\Delta\text{S}_{\text{Neumann}}^{\text{B,C}}$. On the other hand, for $\Delta\text{S}_{\text{Neumann}}^{\text{A}}$ and $\sigma$, their maximum did not coincide at the same energy, making the relation biunivocal: two variations of $\Delta\text{S}_{\text{Neumann}}^{\text{A}}$ corresponded to the same cross section $\sigma$. On the other hand, the variations $\Delta\langle S_z\rangle$ for the state $\ket{\uparrow(\downarrow)}\otimes\Psi^{\pm}(\Phi^{\pm})$ presented a  minimum at $E=1.18m_\mu$, and increased monotonically to zero as $E\rightarrow\infty$. The opposite was for the state $\ket{\downarrow(\uparrow)}\otimes\Psi^{\pm}(\Phi^{\pm})$, where there was a maximum at the same point and decreased monotonically to zero as $E\rightarrow\infty$. The coincidence at $E=1.18\, m_\mu$ of these maxima and minima with the maximum of the cross section allowed to obtain a linear relation (see figure \ref{fig:f11}) which, as the W state, enables us to obtain direct information about the scattering process from spin measurements on the spectator particle $C$.

It is important to notice that if we considered system $A$ initially in a non polarized state, say a statistical mixture $\rho_\text{A} = (2)^{-1}[\ket{\uparrow}\bra{\uparrow} +\ket{\downarrow}\bra{\downarrow}]$ and $BC$ entangled in spin as 
\begin{equation}
	\ket{\mathscr{S}}_{\text{BC}} = d_{1}\ket{\uparrow\uparrow}+d_{2}\ket{\uparrow\downarrow}+d_{3}\ket{\downarrow\uparrow} 
	+d_{4}\ket{\downarrow\downarrow}.
\end{equation}
In \cite{PRD2019}, it was studied the case $d_2=d_3=0$ and $d_1 = \cos \eta$,  $d_2 = e^{i \beta} \sin \eta$ which yields, for the initial state,
$$\rho_{\text{in}}^{\text{ABC}}=
\begin{pmatrix}
\frac{1}{2} & 0\\\\
0 & \frac{1}{2}
\end{pmatrix}\otimes
\begin{pmatrix}
\cos^{2}\eta & 0 & 0 & e^{-i\beta}\sin\eta\cos\eta\\
0 & 0 & 0 & 0\\
0 & 0 & 0 & 0\\
e^{i\beta}\sin\eta\cos\eta & 0 & 0 & \sin^{2}\eta
\end{pmatrix}.$$
Interestingly, with such an initial state as input, before and after the scattering the resulting reduced density matrices of $A$, $B$ and $C$ systems turn out to be identical to the GHZ density matrices seen in equation (\ref{ghzall}). Consequently, for this particular state we would have $\Delta \langle S_{x,z}^{\text{C}} \rangle= 0$, contrarily to the conclusion in \cite{PRD2019}. This result is guaranteed for any other combination of $d_{j}$, since entanglement of two particles implies the same behavior in the individual systems $B$ and $C$ as seen in $\ket{\Psi^\pm(\Phi^\pm)}$. A combination in which all the $d_{j}$ are different from $0$ would lead to a product state in which $C$ could clearly be isolated from the scattering process between $A$ and $B$.

The inelastic collision we have analyzed did not present any pole divergence in the scattering polar angle, since the characteristic amplitude for this process is given by the \textit{s}-channel, $\mathcal{M}\propto[2E^{2}]^{-1}$, whose only restriction is the threshold energy $E=105.7 \text{MeV}$, corresponding to the mass of the muon. This allowed us to obtain a relation with the total cross section $\sigma$, analytical in its entire domain $0\leq\theta\leq\pi$ and $0\leq\phi\leq2\pi$. For the elastic collisions, M\o ller scattering ($e^{-}e^{-}\rightarrow e^{-}e^{-}$) and Bhabha scattering ($e^{-}e^{+}\rightarrow e^{-}e^{+}$), whose amplitudes include the \textit{t} and \textit{u} channels, $\mathcal{M}\propto[-2p^{2}(1\pm\cos\theta)]^{-1}$, a different behavior would be expected when including the non-analytic variable $\theta$ in its entire domain ($\theta=0,\pi$). In this case, the spin density matrices can not be integrated over the variable $\theta$, and it is more appropriate to establish a relation with the differential cross section $d\sigma/d\theta$.

Finally it is important to recall that we have specifically calculated the contribution to the spin expectation value of particle $C$ stemming from the muonic channel of electron positron scattering. For the range of energy that we studied, photon production takes place. The total cross section for this process grows monotonically with the energy whereas the cross section for muonic channel has a peak at $E=1.2\, m_{\mu}$. Because there are no interference terms between these channels, we may in principle separate their contributions in $\Delta\braket{S_{i}^C}$. Other channels such as $e^{+}e^{-}\rightarrow e^{+}e^{-}$ and hadron production (e.g. $e^+ e^- \rightarrow$ pions) are dominant far from the range of energy $m_\mu \rightarrow 5 \, m_\mu$, taking place at much lower and higher energies, respectively \cite{JEGER, GROOTE, PESKIN}.

Although an explicit computation of the cross section for all the competing processes in the range of energy
where muon production is predominant allows  numerical evaluation of its relative  contribution to the theoretical value $\Delta \langle S_i^C \rangle$, post-selective measurements of muon decays are statistically more judicious \cite{ABL, NORI}.  In such framework, the spin of the spectator particle is measured after scattering followed by the post-selection by means of identifying and selecting the muonic scattering channel (subensemble) for the statistics. A similar analysis was used in  Ref.\cite{MEDINA}  for the analysis of entanglement enhancement of a two-electron inelastic scattering  and in Ref. \cite{FEDER} for the study  of entanglement and entropy in electron-electron scattering.

As future perspectives of work, it would be interesting to study the effect of Lorentz boosts on entanglement of the tripartite state considered in this contribution in the same sense as discussed in Ref. \cite{BERTLMANN} but bearing in mind that relativistic observables are in general observer dependent. In other words, distinct inertial observers measure the same value for a certain
observable provided the experiments are carried on with
states equally prepared in the corresponding proper frames \cite{MATSAS}. Another interesting venue to be analysed is to study dressed final states to account for soft radiation and information leak in the process in order to assess how much information is effectively avaliable \cite{CARNEY}.

\begin{acknowledgments}
MS thanks CNPq for a research grant 302790/2020-9. IGP thanks CNPq for the Grant No.307942/2019-8. We acknowledge support from CFisUC and FCT through the project UID/FIS/04564/ 2020, grant CERN/FIS-COM/0035/2019. JDF was financed in part by the Coordenação de Aperfeiçoamento de Pessoal de Nível Superior - Brasil (CAPES).

\end{acknowledgments}


\begin{thebibliography}{99}
\bibitem{EPR} A. Einstein, B. Podolsky and N. Rosen, Phys. Rev. 47 (1935) 777.
\bibitem{BELL} J. S. Bell, Physics 1 (1964) 1964.
\bibitem{GRANGIER} P. Grangier, quant-ph/2012.09736.
\bibitem{ASPECT} A. Aspect, P. Grangier and G. Roger, Phys. Rev. Lett. 47 (1981) 460; A. Aspect, J. Dalibard, G. Roger, Phys. Rev. Lett. 49 (1982) 1804.
\bibitem{LOOPHOLE1} B. Hensen et al., Nature 526 (2015) 682.
\bibitem{LOOPHOLE2} W. Rosenfeld et al., Phys. Rev. Lett. 119 (2017) 010402.
\bibitem{RQIREVIEW} D. R. Terno, ``Quantum Information Processing: From theory to experiment'', edited by D. G. Angelakis et al., IOP Press (2006).
\bibitem{FRIIS} N. Friis, A. R. Lee and J. Louko, Phys. Rev. D 88 (2013) 064028; N. Friis, M. Huber, I. Fuentes and D. E. Bruschi, Phys. Rev. D 86 (2012) 105003; N. Friis, A. R. Lee, D. E. Bruschi and J. Louko, Phys. Rev. D 85 (2012) 025012.
\bibitem{BERTLMANN} N. Friis, R. A. Bertlmann, M. Huber and B. C. Hiesmayr, Phys. Rev. A 81 (2010) 042114.
\bibitem{AHN} D. Ahn, H.-J. Lee, Y. H. Moon and S. W. Hwang, Phys. Rev. A 67 (2003) 012103.
\bibitem{RQIIF} D. E. Bruschi, A. R. Lee and I. Fuentes, J. Phys. A: Math. Theor, 46 (2013) 165303; A. R. Lee and I. Fuentes, Phys. Rev. D 89 (2014) 085041.

\bibitem{GW} F. Khalili, E. S. Polzik, Phys. Rev. Lett. 121 (2018) 031101.
\bibitem{PID} Lukasz Dusanowski, S.-H. Kwon, C. Schneider and S. H\"{o}fling, Phys. Rev. Lett. 122 (2019) 173602.
\bibitem{BALL} J. L. Ball, I. F.-Schuller, F. P. Schuller, Phys. Lett. A 359 (2006) 550.
\bibitem{ALICE} I. F.-Sch\"uller and R. B. Mann, Phys. Rev. Lett. 95 (2005) 120404.
\bibitem{LOUKO} N. Friis, D. E. Bruschi, J. Louko and I. Fuentes, Phys Rev. D 85 (2012) 081701; N. Friis, P. K\"{o}hler, E. M. Martinez and R. A. Bertlmann, Phys. Rev. A 84 (2011) 062111.
\bibitem{FEDER} P. Schattschneider, S. L\"{o}ffler, H. Gollish and R. Feder, J. El. Spect. And Related Phen. 241 (2020) 146810.
\bibitem{GHANBARI} Ebrahim G.-Adivi and M. Soltani, Eur. Phys. J. D 68 (2014) 336.
\bibitem{HARSHMAN4} N. L. Harshman and S. Wickramasekara, Phys. Rev. Lett. 98 (2007) 080406.
\bibitem{HARSHMAN1} N. L. Harshman, Int. J. Mod. Phys. A 20 (2005) 6220.
\bibitem{HARSHMAN2} N. L. Harshman, Phys. Rev. A 73 (2006) 062326.
\bibitem{HARSHMAN3} N. L. Harshman, Int. J. Q. Inf. 5 (2007) 273.
\bibitem{HARSHMAN5} N. L. Harshman and P. Singh, J. Phys. A: Math. Theor. 41 (2008) 155304.
\bibitem{LAMATALEON} L. Lamata and J. L\'eon, Phys. Rev. A 73 (2006) 052322.
\bibitem{KOUZAKOV} K. A. Kouzakov, Theor. Math. Phys. 201 (2019) 1664.
\bibitem{PACHOS} J. Pachos and E. Solano, QIC 3 (2003) 115.
\bibitem{MISHIMA} K. Mishima, M. Hayashi and S. H. Lin, Phys. Lett. A 333 (2004) 371.
\bibitem{LAMATASOLANO} L. Lamata, J. L\'eon and E. Solano, Phys. Rev. A 73 (2006) 012335.
\bibitem{WANG} H.-J. Wang and W. T. Geng, J. Phys. A: Math. Theor. 40 (2007) 11617.
\bibitem{BUSCEMI} F. Buscemi, P. Bordone and A. Bertoni, Phys. Rev. A 75 (2007) 032301.

\bibitem{102001} S. R. Beane, D. B. Kaplan, N. Klco, and M. J. Savage, Phys. Rev. Lett. 122, 102001 (2019).
\bibitem{168581} S R. Beane, R. C. Farrell, Ann. Phys. 433 (2021) 168581.
\bibitem{2150205} S. R. Beane, R C. Farrell, and M. Varma, Int.J.Mod.Phys.A 36 (2021) 30, 2150205.


\bibitem{SEKI1} R. Peschanski and S. Seki, Phys. Lett. B 758 (2016) 89.
\bibitem{SEKI2} R. Peschanski and S. Seki, Phys. Rev. D 100 (2019) 076012.
\bibitem{SEKI3} S. Seki, I. Y. Park and S.-J. Sin, Phys. Lett. B 743 (2015) 147.
\bibitem{Faleiro} R. Faleiro, R. Pavão, H. A. S. Costa3, B. Hiller, A. H. Blin and M. Sampaio, J. Phys. A 53 (2020)365301.
\bibitem{FAN1} J. Fan, Y. Deng and Y.-C. Huang, Phys. Rev. D 95 (2017) 065017.
\bibitem{FAN2} J. Fan, X. Li, Phys. Rev. D 97 (2018) 016011.
\bibitem{YONGRAM} N. Yongram and E. B. Manoukian, Fortschr. Phys. 61 (2013) 668.
\bibitem{INADA} T. Inada et al., Phys. Lett. B 732 (2014) 356.
\bibitem{RATZELQED} D. R\"{a}tzel, M. Wilkens and R. Menzel, Phys. Rev. A 95 (2017) 012101.
\bibitem{RATZELQG} D. R\"{a}tzel, M. Wilkens and R. Menzel, Europhys. Lett. 115 (2016) 51002.
\bibitem{CABAN1} M. Wlodarczyk, P. Caban, J. Ciborowski, M. Dragowski and J. Rembielinski, Phys. Rev. A 95 (2017) 022103.
\bibitem{CABAN2} P. Caban,  J. Rembielinski, M. Wlodarczyk, , J. Ciborowski, M. Dragowski and A. Poliszuk, Eur. Phys. J Web of Conferences 164 (2017) 07031.
\bibitem{TSURIKOV} D. E. Tsurikov, S. N. Samarin, J. F. Williams and O. M. Artamonov, J. Phys. B: At. Mol. Opt. 50 (2017) 075502.
\bibitem{KLN} T. Kinoshita, J. Math. Phys. 3 (1962) 650; T. D. Lee and Nauenberg, Phys. Rev. 113 (1964) 1549.
\bibitem{GOMEZ} C. Gomez, R. Letschka and S. Zell, Eur. Phys. J. C 78 (2018) 610.
\bibitem{CARNEY} D. Carney, L. Chaurette, D. Neuenfeld and G. W. Semenoff, Phys. Rev. Lett. 119 (2017) 180502; idem, JHEP 09 (2018) 121.
\bibitem{TOMARAS} T. N. Tomaras and N. Toumbas, Phys. Rev. D 101 (2020) 065006.

\bibitem{PRD2019} J. B. Araujo, B. Hiller, I. G. Paz, M. M. Ferreira Jr., M. Sampaio and H. A. S. Costa, Phys. Rev. D. 100 (2019) 105018.

\bibitem{vidal} W. D\"ur, G. Vidal and J. I. Cirac,  Phys.  Rev. A 62 (2000) 062314.

\bibitem{CRUZ} D. Cruz, R. Fournier, F. Gremion, A. Jeannerot, K. Komagata, T. Tpsic, J. Thiesbrummel, C. L. Chan, N. Macris, M.-Andr\'e Dupertuis, C. J.-Galy, Adv. Quant. Technol. (2019) 1900015.




\bibitem{JUNG} Eylee Jung, Mi-Ra Hwang, You Hwan Ju, Min-Soo Kim, Sahng-Kyoon Yoo, Hungsoo Kim, D. K. Park, Jin-Woo Son, S. Tamaryan and Seong-Keuck Cha, Phys. Rev. A 78 (2008) 012312.


\bibitem{horodick} R. Horodecki, P.Horodecki, M. Horodecki, and K. Horodecki, K. (2009). Quantum entanglement. Reviews of modern physics, 81(2), 865.

\bibitem{bellstates} S. L. Braunstein, A. Mann and M. Revzen, Phys. Rev. Lett. 68 (1992) 3259.

\bibitem{MATSAS} L. H. Zambianco, A. G. S. Landulfo and G. Matsas, Phys. Rev. A 1000 (2019) 062126.

\bibitem{JEGER} F. Jegerlehner and  K. Kolodziej, Eur. Phys. J. C 77 (2017) 254.

\bibitem{GROOTE} S. Groote, The electron-positron annihilation cross section used
for high precision tests of the Standard Model, MZ-TH/02-30, hep-ph/0212041.

\bibitem{PESKIN} M. E. Peskin and D. V. Schroeder, An Introduction to Quantum Field Theory,  CRC Press (2018).

\bibitem{ABL} Y. Aharonov, P. G. Bergmann, J. L. Lebowitz,  Phys. Rev. 134 (1964) 1410.

\bibitem{NORI} A. G. Kofman, S. Ashhab and F. Nori, Phys. Rep. 520 (2012) 43.

\bibitem{MEDINA} A. Lopez, V. M. Villalba and E. Medina, Phys. Rev. B 76 (2007) 115107.

\bibitem{FEDER} P. Schattschneider, S. L\"offler,  H. Gollisch and R. Feder, Journal of Electron Spectroscopy and Related Phenomena 241 (2020) 146810.






















\end{thebibliography}
\end{document}